\documentclass[12pt]{article}
\usepackage{amsmath, amsthm, amscd, amsfonts, amssymb, graphicx, color}
\usepackage{graphics,epsfig}
\oddsidemargin 10mm \numberwithin{equation}{section}
\def\be{\begin{equation}}
\def\ee{\end{equation}}
\begin{document}
\begin{center}
{\bf {Gravitational lensing of charged Ayon-Beato-Garcia black
holes and non-linear effects of Maxwell fields }}
 \vskip 0.50 cm
  {H. Ghaffarnejad\footnote{E-mail address:
   hghafarnejad@semnan.ac.ir
}, M. A. Mojahedi\footnote{E-mail address:
amirmojahed@semnan.ac.ir} and H. Niad \footnote{E-mail address:
niad@semnan.ac.ir} }\vskip 0.1 cm \textit{Faculty of Physics,
Semnan University, Zip Code 35131-19111, Iran}
\end{center}
 \begin{abstract} Non-singular Ayon-Beato-Garcia
(ABG) spherically symmetric sta\\tic black hole (BH) with charge
to mass ratio $q=\frac{g}{2m}$  is metric solution of Born-Infeld
nonlinear Maxwell-Einstein theory. Central region of the BH
behaves as (anti-) de Sitter  for $(|q|>1)~|q|<1 .$ In case of
$|q|=1$ the BH central region behaves as Minkowski flat metric.
Nonlinear Electromagnetic (NEM) fields counterpart causes to
deviate light geodesics and so light rays will forced to move on
effective metric. In this paper we study weak  and strong
gravitational lensing of light rays by seeking affects of NEM
fields counterpart on image locations and corresponding
magnification. We set our calculations to experimentally observed
Sgr A$^*$  BH. In short we obtained: For large distances the NEM
counterpart is negligible reaching to linear Maxwell fields. The
NEM makes enlarge the BH photon sphere radius as linearly by
raising $|q|>1$ but deceases by raising $|q|\leq1.$ Sign of
deflection angle of bending light rays is changed in presence of
NEM effects with respect to ones obtained in absence of NEM
fields. Absolute value of deflection angle raises by increasing
$|q|\to1.$ Weak image locations decreases (increases) by raising
$0<|q|<1$ in presence (absence) of NEM fields. By raising the
closest distance of the bending light rays weak image locations
changes from left (right) to right (left) in absence (presence) of
NEM fields. Einstein rings radius and corresponding magnification
centroid  become larger (smaller) in presence (absence) of NEM
fields in case of weak lensing. Angular separation $s$ between the
innermost and outermost relativistic images increases (decreases)
by increasing $0<|q|<1$ in absence (presence) of NEM fields.
Corresponding magnification
 $r$ decreases (increases) by raising $0<|q|<1$ in absence (presence)
of NEM fields. $s (r)$ raises (decreases) by increasing $|q|>>1.$
\end{abstract}
\section{Introduction}
Since the advent of Einstein's general relativity theory, black
holes and the singularity problem of curved space times become
challenging subjects in modern physics because of presence of
quantum physics. Singularity is the intrinsic character of the
most exact solutions of Einstein's equations where Ricci and
Kretschmann scalars reach to infinite value at singular point of
the space time  [1]. Penrose cosmic censorship conjecture states
that the causal singularities must be covered by the event horizon
and so causes to disconnect interior and exterior regions of the
space time [2,3]. However nonsingular metric solutions are also
obtained from the Einstein field equation (see for instance
[4-23]). In the latter situations the Einstein field equation is
coupled to suitable NEM fields for which the Ricci and the
Kretschmann scalars become regular in whole space time. A good
classification of spherically symmetric static regular black holes
are collected in ref. [9]. Inspiring a physical central core idea,
Bardeen suggested the first spherically symmetric static regular
black hole in 1968 containing a horizon without singularity [10].
After his work, other regular black holes were designed based on
this model which we call here for instance ABG [11-14], Hayward
(HAY) [15] and Neves-Saa (NS) [16,17]. Non-singular property of
all of these solutions are controlled via dimensionless charge
parameter $q.$ HAY type of regular black hole is obtained by
modifying the mass parameter of the BAR black hole. NS type of
regular black hole is a HAY type but its asymptotic behavior
approaches to a vacuum de Sitter in presence of cosmological
constant parameter. Regular black holes are studied also on brane
words (see [17] and reference therein). The solutions of rotating
regular black holes have been introduced in several articles
[18-24]. A very important source of strong gravity is the
Kerr-Newman-de Sitter (KNDS) and/or Kerr-Newman- anti-de Sitter
(KNADS) black hole. Kraniotis studied gravitational lensing of
KNDS and KNADS black hole in ref. [25], where closed form analytic
solutions of the null geodesics and the gravitational lens
equations have been obtained in terms of Appell-Lauricella
generalized hypergeometric functions and the elliptic functions of
Weierstrass. In these exact solutions all the fundamental
parameters of the theory, namely black hole mass, electric charge,
rotation angular momentum and the cosmological constant enter on
an equal footing while the electric charge effect on relativistic
observable was also investigated. Rotating nonsingular black holes
can be treat as natural particle accelerators [24]. Ultra-high
energy particle collisions are studied on the regular black holes
[26] and backgrounds containing naked singularity [27]. Motion of
test particles is studied in regular black hole space–times in
ref. [28]. Circular geodesics are obtained for BAR and ABG regular
black-holes in ref. [29]. The optical effects related to Keplerian
discs orbiting Kehagias-Sfetsos (KS) naked singularities was
investigated in ref. [30]. Authors of the latter work are also
mentioned to be close similarity between circular geodesics in KS
and properties of the circular geodesics of the RN naked singular
space times. Schee et al studied also profiled spectral lines
generated by keplerian discs orbiting in the Bardeen and ABG space
times in ref. [31]. Correspondence between the black holes and the
FRW geometries are studied for non-relativistic gravity models in
ref. [32]. RN black hole gravitational lensing is studied in ref.
[33]. Gravitational lensing from regular black holes is studied in
weak deflection limits of light rays [34-36] and in strong
deflection limits of light rays [37-41]. Strong deflection limits
of light rays can be distinguish gravitational lensing between
naked singularity and regular black holes background [41].
 There is significant difference between optical
phenomena characters of the singular space-times such as SCH, RSN,
and non-singular space-times as HAY, BAR, ABG [38]. It is related
to the fact that the regular space-times reach to a de Sitter
and/or anti-de Sitter like approximately at center $r\to0$ (see
Eqs. (2.7) and (2.9)). Furthermore we should point that the
nonsingular charged black holes obtained from  NEM models in
curved space times cause that the photons do not move along null
geodesics. As an applicable approach we must be obtain
corresponding effective metric for geodesics of moving photons
[42-45] and so study their gravitational lensing. The black hole
electric charge has also important effects on final state of the
Hawking radiation and switching off effects of a quantum
evaporating black hole (see for instance [46]). In this work we
study gravitational lensing of light rays moving on the ABG
nonsingular black hole in presence of NEM fields counterparts. The
paper is organized as
follows.\\
Briefly, we introduce in section 2 regular ABG black hole metric
and its asymptotically behavior against different values of $q$.
In section 3 we calculate effective metric of the ABG black hole
for the moving photons by regarding the results of the original
work [12]. We solve numerically the photon sphere equation of
effective metric and obtain photon sphere radius against different
charge values $q.$ In section 4 we evaluate general formalism of
deflection angle of bending light rays in weak and strong
deflection limits. In weak deflection limits we apply the Ohanian
lens equation [47] to determine non-relativistic image locations
against source positions for observed Sgr $A^*$ black hole
[48-51]. Weak deflection angle of bending light rays and their
magnifications are evaluated numerically point by point and they
are plotted against source locations and also $q$. In the strong
deflection limits we use Bozza`s formalism [37,38] to obtain
logarithmic form of the deflection angle. We obtain relative
distance between innermost and outermost relativistic images and
corresponding magnification and then plot their diagrams.
 Section 5
denotes to concluding remark.
\section{ABG space time}
The ABG spherically symmetric black hole metric defined by
Schwarzschild coordinates is [12] \be
ds^2=-H(r)dt^2+\frac{dr^2}{H(r)}+r^2(d\theta^2+\sin^2\theta
d\varphi^2)\label{2.1}\ee with \be H(r)=1-\frac{2m
r^2}{\big(r^2+g^2\big)^{3/2}}+\frac{g^2
r^2}{\big(r^2+g^2\big)^{2}}\ee and associated electric field \be
F_{tr}(r)=E(r)=
gr^4\bigg(\frac{r^2-5g^2}{(r^2+g^2)^4}+\frac{15}{2}\frac{m}{(r^2+g^2)^{\frac{7}{2}}}\bigg).\ee
$m$ and $g$ are total mass and electric charge parameters of the
BH respectively. The line element (2.1) is non-singular static
solution of NEM-Einstein metric equation \be G_{\mu\nu}=8\pi
T_{\mu\nu}=8\pi\{\mathcal{L}_FF_{\mu\eta}F^{\eta}_{\nu}-\mathcal{L}g_{\mu\nu}\},~~~\mathcal{L}_F=\frac{\partial\mathcal{L}}{\partial
F}\ee which satisfies  the action functional $ I=\int\sqrt{g}
dx^4\big(\frac{R}{16\pi}-\frac{\mathcal{L}(F)}{4\pi}\big)$ where
$R$ is Ricci scalar and $\mathcal{L}$  is a functional  of
$F=\frac{1}{4}F_{\mu\nu}F^{\mu\nu}.$ This metric solution has only
the coordinate singularity called as horizon singularity because
the Ricci and the Kretschmann scalars become regular at all points
of the space time $0\leq r\leq+\infty$. Defining mass and charge
functions as
\begin{equation} M(r)=m\bigg(1+\frac{g^2}{r^2}\bigg)^{-\frac{3}{2}},~~~~e(r)=g\bigg(1+\frac{g^2}{r^2}\bigg)^{-1}\end{equation}
one can show that the ABG metric (2.1) reduces apparently  to a
variable mass-charge RN type of BH as
\begin{equation}ds^2=-\bigg(1-\frac{2M(r)}{r}+\frac{e^2(r)}{r^2}\bigg)dt^2+\frac{dr^2}{\bigg(1-\frac{2M(r)}{r}+\frac{e^2(r)}{r^2}
\bigg)}+r^2(d\theta^2+\sin^2\theta d\varphi^2)\end{equation} where
$M(\infty)=m$ and $e^2(\infty)=g$ are  ADM mass and electric
charge viewed from observer located at infinity.  Its central
region  $0<r<|g|$ behaves as vacuum de Sitter asymptotically:
\begin{equation}ds^2\approx-\bigg(1-\frac{\Lambda}{3}r^2\bigg)dt^2+\frac{dr^2}
{\bigg(1-\frac{\Lambda}{3}r^2\bigg)}+r^2(d\theta^2+\sin^2\theta
d\varphi^2)\end{equation} for \begin{equation}
|q|=\frac{g}{2m}<1\end{equation} and anti de Sitter
\begin{equation}ds^2\approx-\bigg(1+\frac{\Lambda}{3}r^2\bigg)dt^2+
\frac{dr^2}{\bigg(1+\frac{\Lambda}{3}r^2\bigg)}+r^2(d\theta^2+\sin^2\theta
d\varphi^2)\end{equation} for \begin{equation}
|q|=\frac{g}{2m}>1\end{equation} respectively where we defined
effective cosmological constant as
\begin{equation}\Lambda(m,g)=\frac{3(1-q)}{4m^2q^3}.\end{equation} In particular case \begin{equation}
|q|=\frac{g}{2m}=1\end{equation} the effective cosmological
parameter vanishes $\Lambda=0$ and so near the center $r\to0$, the
 ABG black hole metric  reduces to a flat Minkowski background
 asymptotically. Setting $g=0$ the equations (2.5) read
 $m=M,e=0$ for which the metric solution (2.1) leads to singular charge-less Schwarzschild BH.  Nonlinear counterpart of the Maxwell stress
 tensor causes to deviate the photon geodesics where the photons
 do not move along the null geodesics. Usually one use an
 effective metric to study gravitational lensing of the light rays
 moving on such a charged black holes  metric [41-44].
  In the following section we seek effective metric of the ABG black hole for photon trajectories.
 \section{Effective metric for photon trajectories}
 Assuming $\mathcal{L}(F)=F,$ the equation (2.4) leads to the well known
 linear Einstein-Maxwell gravity where the photon propagates by the null equation
 \be g_{\mu\nu}k^{\mu}k^{\nu}=0\ee where $k^{\mu}$ is corresponding four-momentum of the photon, but in general form where
  $\mathcal{L}(F)\neq F$ the electric field given by  (2.3), is
 self-interacting and so directly is reflected on the photon
 propagation. In the latter case the photons do not move along null
 geodesics (3.1) but instead, photons propagate along null geodesics of an
 effective geometry which depends on used nonlinear theories
[43,44,52] as
 \be g^{eff}_{\mu\nu}k^{\mu}k^{\nu}=0\ee
where \be
g_{\mu\nu}^{eff}=16\bigg(\frac{\mathcal{L}_{FF}F_{\mu\eta}F^{\eta}_{\nu}-(\mathcal{L}_F+2F\mathcal{L}_{FF})
g_{\mu\nu}}{F^2\mathcal{L}^2_{FF}-16(\mathcal{L}_F+F\mathcal{L}_{FF})^2}\bigg)\ee
and \be
g^{\mu\nu}_{eff}=\mathcal{L}_{FF}F^{\mu}_{\eta}F^{\nu\eta}+\mathcal{L}_F
g^{\mu\nu}.\ee  In absence of nonlinear counterpart of EM fields
we must be set \be
\mathcal{L}_{FF}=0,~~~\mathcal{L}_F=1,~~~\mathcal{L}=F\ee for
which the effective metric reaches $g_{\mu\nu}^{eff}\to
g_{\mu\nu}.$  We are now in position to obtain effective metric of
spherically symmetric static space time (2.1). To do so  we must
be obtain all quantities defined by
$\{\mathcal{L},\mathcal{L}_F,\mathcal{L}_{FF}\}$ which satisfy the
metric solution (2.1). We should first obtain corresponding
Lagrangian density $\mathcal{L}(F).$ We use result of the original
paper [12] where its authors are used following ansatz to solve
(2.4) and obtain (2.1). \be
F\mathcal{L}_F^2=-\frac{1}{2(2m)^2}\frac{q^2}{x^4}\ee  where \be
F(x)=-\frac{1}{(2m)^2}\frac{q^2x^8}{2}\bigg[\frac{x^2-5q^2}{(x^2+q^2)^4}+\frac{15}{4}\frac{1}{(x^2+q^2)^{\frac{7}{2}}}\bigg]^2\ee
comes from (2.3) by inserting dimensionless electric charge $q$
and radial coordinate $x$ \be q=\frac{g}{2m},~~~x=\frac{r}{2m}\ee
into $F=\frac{1}{4}F_{\mu\nu}F^{\mu\nu}.$ One can obtain
asymptotically behavior of the equation (3.7) for large distances
$x>>1$ as $F_{\infty}(x)\approx-\frac{1}{2(2m)^2}\frac{q^2}{x^4}.$
Comparing the latter result and (3.6) we infer
$\mathcal{L}_{F}\approx1$ which by integrating leads to linear
Maxwell Lagrangian $\mathcal{L}\to F.$ The latter result tells  us
NEM action functionals $\mathcal{L}(F)$ are negligible for regions
of far from the black hole event horizon $x>>x_{EH}.$ Applying
(3.6) and (3.7) we  obtain parametric form of the Lagrangian
density $\mathcal{L}(x)$ as follows. \be
\mathcal{L}(x)=-\frac{q^2}{(2m)^2}\int_{\infty}^x\frac{1}{x^2}d\bigg\{
x^4\bigg[\frac{x^2-5q^2}{(x^2+q^2)^4}+\frac{15}{4}\frac{1}{(x^2+q^2)^{\frac{7}{2}}}\bigg]\bigg\}\ee
which has exact solution as \be
\mathcal{L}(x)=-\frac{q^2}{(2m)^2}\bigg[\frac{x^2(x^2-5q^2)}{(x^2+q^2)^4}+\frac{15}{4}\frac{x^2}{(x^2+q^2)^{\frac{7}{2}}}\bigg]
+$$$$\frac{q^2}{(2m)^2}\bigg[\frac{1}{2(x^2+q^2)^2}-\frac{2q^2}{(x^2+q^2)^3}+\frac{3}{2}\frac{1}{(x^2+q^2)^{\frac{5}{2}}}\bigg].\ee
One infers \be
\mathcal{L}_F(x)=\frac{\mathcal{L}^{\prime}(x)}{F^{\prime}(x)}\ee
and \be
\mathcal{L}_{FF}(x)=\frac{1}{F'(x)}\bigg(\frac{\mathcal{L}'}{F'}\bigg)'=\frac{\mathcal{L}''F'-F''\mathcal{L}'}{{F'}^3}\ee
where over-prime  $\prime$ denotes to differentiation with respect
to $x.$ If we need to obtain exact form of the functional
$\mathcal{L}(F)$ we must be remove $x$ between (3.7) and (3.10)
but it will take more complex form. Hence we plot numerical
diagram of $\mathcal{L}(F)$ by inserting numerical values of
tables 1 and 2 in figure 1. The diagram shows that negligibility
of NEM fields for $|q|>1$ but not for $|q|<1.$  However we will
need to exact form of the functions
$\{\mathcal{L}(x),\mathcal{L}_F(x),\mathcal{L}_{FF}(x)\}$ to study
location of effective metric horizons, gravitational lensing
images and their magnifications. To do so we will use numerical
method as follows.  For metric solution (2.1) one can show that
the effective metric (3.3) become \be
ds^2_{eff}=-A(r)dt^2+B(r)dr^2+r^2C(r)(d\theta^2+\sin^2\theta
d\varphi^2)\ee where we defined \be
A(r)=\frac{16H(r)\mathcal{L}_F}{16(\mathcal{L}_F+F\mathcal{L}_{FF})^2-F^2\mathcal{L}_{FF}^2}\ee
\be
B(r)=\frac{1}{H(r)}\frac{16\mathcal{L}_F}{16(\mathcal{L}_F+F\mathcal{L}_{FF})^2-F^2\mathcal{L}_{FF}^2}\ee
and \be
C(r)=\frac{8(2\mathcal{L}_F+4F\mathcal{L}_{FF})}{16(\mathcal{L}_F+F\mathcal{L}_{FF})^2-F^2\mathcal{L}_{FF}^2}.\ee
 The radius of the event horizon $r_{H}$ is given by the greatest
positive root of the equation $H(r)=0(A(r)=0)$ in absence
(presence) of nonlinear counterpart of EM field. According to
study of black hole gravitational lensing, photon sphere
construction is one of important characters
  which must be considered here. It comes from energy condition
[53] and is a particular hyper-surface ($r=constant$) which does
not evolve with time. In other words any null geodesic initially
tangent to the photon sphere hyper-surface will remain tangent to
it. It is made from circulating photons turn turning around the
black hole center. Radius of the photon sphere $r_{ps}$ is the
greatest positive solution of the equation
 [48] \be \bigg(
 \frac{1}{r^2}\frac{A(r)}{C(r)}\bigg)_{|_{r=r^{eff}_{ps}}}^{\prime}=0.\ee
  Setting (3.5)
the equations (3.14), (3.15)  and (3.16) read \be A(r)=H(r),
~~~B(r)=\frac{1}{H(r)},~~~C(r)=1\ee describing original space time
(2.1) in absence of the nonlinear EM fields effects for which
(3.17) become \be
\bigg(\frac{H(r)}{r^2}\bigg)^{\prime}_{|_{r=r_{ps}}}=0.\ee
Diagrams of the equations (3.17) and (3.19) are plotted for larger
solutions  in figure 1.  Linear branch of the right panel of
diagram in figure 1 predicts  large scale photon spheres for
$|q|>1$ which are formed only in presence of NEM field. This
linear branch of the effective photon sphere diagram can be
approximated with the following equation. \be
x_{ps}^{eff}(|q|>1)\approx3.643|q|-0.796\ee which raises by
increasing $|q|\to\infty.$ We calculated numerical values of the
above photon sphere radius for $1<|q|<36$ and collected in the
table 2. Corresponding diagram is given in figure 1.  One can
result from the figure 1 that we have small scale photon sphere
for $|q|<1$ from both of the effective metric (3.13) and the
original one (2.1). Hence obtained gravitational lensing results
from (2.1) can be compared with ones which obtained from (3.13)
only for $|q|<1.$ Thus we collect numerical solutions of the both
photon sphere equations (3.17) and (3.19) for $|q|<1$ in table 1.
We will need them to evaluate numerical values of deflection
angle, image locations and corresponding magnifications.  We will
study gravitational lensing of the system separately for two
regimes $|q|>1$ and $|q|<1$ as follows. We first apply to evaluate
numerical values of the deflection angle of bending light rays.
\section{Deflection angle} When light ray moves at
neighborhood of the ABG black hole and deflects without turning
around the black hole center then gravitational lensing takes
`weak deflection limits` approach. In the latter case closest
approach distance of the bending light rays from the black hole
center $r_0$ become larger than the photon sphere radius and two
non-relativistic images are usually formed. They are called as
primary and secondary images. In general, bending angle of light
rays is obtained by solving null geodesics equation defined by
(3.1) as follows [54]. \be\alpha_{eff}(r_0)=I_{eff}(r_0)-\pi\ee
where \be
I_{eff}(x_0)=2\int_{x_0>x_{ps}}^{\infty}\frac{\sqrt{A(x)B(x)/C^2(x)}}{\sqrt{\frac{A(x_0)}{x_0^2C(x_0)}-\frac{A(x)}{x^2C(x)}}}\frac{dx}{x^2}.
\ee  Inserting \be z=\frac{x_0}{x}\ee the integral equations (4.2)
become  \be
I_{eff}(x_0)=2\int_0^1\frac{\Gamma(\frac{x_0}{z})}{\sqrt{\Omega(x_0)-\Omega(\frac{x_0}{z})z^2}}dz\ee
where we defined \be
\Gamma\big(\frac{x_0}{z}\big)=\Omega\sqrt{\frac{B}{A}}=
\frac{\mathcal{L}_F}{\mathcal{L}_F+2F\mathcal{L}_{FF}},~~~\Omega\big(\frac{x_0}{z}\big)=\frac{A}{C}=H\Gamma\ee
According to method given in ref. [52], we now expand
$\Gamma(\frac{x_0}{z})$ and $\Omega(x_0)-\Omega(\frac{x_0}{z})z^2$
in powers of $(1-z)$ as follows. \be
\Gamma\big(\frac{x_0}{z}\big)=\Gamma_0+\Gamma_1(1-z)+\Gamma_2(1-z)^2+O(3)\ee
and \be
\Omega(x_0)-\Omega\big(\frac{x_0}{z}\big)z^2=\Omega_1(1-z)+\Omega_2(1-z)^2+O(3)\ee
where we defined \be
\Gamma_0=\Gamma(x_0),~~~\Gamma_1=x_0\Gamma^{\prime}(x_0),~~~\Gamma_2=x_0\Gamma^{\prime}(x_0)+x_0^2\Gamma^{\prime\prime}(x_0)/2\ee
and \be
\Omega_1=2\Omega(x_0)-x_0\Omega^{\prime}(x_0),~~~~\Omega_2=x_0\Omega^{\prime}(x_0)-\Omega(x_0)-x_0^2\Omega^{\prime\prime}(x_0)/2\ee
in which over-prime $^\prime$ denotes to differentiation with
respect to its  argument $x.$ Inserting (4.6) and (4.7) and
neglecting their higher order terms, the integral equation (4.4)
 become \be
I_{eff}(x_0)\approx2\int_0^1dz\bigg[\frac{\Gamma_0+\Gamma_1(1-z)+\Gamma_2(1-z)^2}{\sqrt{\Omega_1(1-z)+\Omega_2(1-z)^2}}\bigg]\ee
which  has  solution as follows.
 \be I_{eff}(x_0)=\frac{1}{\sqrt{\Omega_2(x_0)}}\sqrt{1+\frac{\Omega_1(x_0)}{\Omega_2(x_0)}}
 \bigg[2\Gamma_1(x_0)+\Gamma_2(x_0)-\frac{3}{2}\Gamma_2(x_0)\frac{\Omega_1(x_0)}{\Omega_2(x_0)}\bigg]$$$$-
 \frac{1}{\sqrt{\Omega_2(x_0)}}\bigg[2\Gamma_0(x_0)-\Gamma_1(x_0)\frac{\Omega_1(x_0)}{\Omega_2(x_0)}+
 \frac{3}{4}\Gamma_2(x_0)\bigg(\frac{\Omega_1(x_0)}{\Omega_2(x_0)}
 \bigg)^2\bigg]$$$$\times\ln\bigg[1+2\
 frac{\big(1-\sqrt{1+\frac{\Omega_1(x_0)}{\Omega_2(x_0)}}\big)}{\frac{\Omega_1(x_0)}{\Omega_2(x_0)}}\bigg]\ee
 Weak (strong) deflection limits of bending light rays are regimes where $I_{eff}(x_0)$ $\nrightarrow\infty(\to\infty).$
 This restrict us to choose particular regimes of the ratio  $\frac{\Omega_1}{\Omega_2}$ given by (4.11).
 Inserting (4.5) and (4.9) into the
 photon sphere
equation (3.17) and setting $x_0=x_{ps}^{eff}$ one can result
$x_{ps}\Omega^{\prime}(x_{ps})-2\Omega(x_{ps})=\Omega_1(x_{ps})=0$.
The latter condition is valid for moving light rays
 near the photon sphere for which  $I_{eff}\to\infty.$ In other words one infers $\Omega_1(x_0\neq x_{ps}^{eff})\neq0$ for weak deflection
 limits and so we can use asymptotic expansion form of the integral solution
(4.11) for $x_0>x_{ps}$ and $\forall q$ as follows.
\subsection{Weak lensing deflection angles} One can obtain asymptotic expansion series form of the functions
 $\Omega_{1,2}(x_0)$ and $\Gamma_{0,1,2}(x_0)$ which up to terms in order
of $O(x_0^{-3})$ become respectively
\be\Omega_1(x_0)\approx2-\frac{\frac{3}{8}}{x_0}-\frac{(32q^2+\frac{345}{8})}{x_0^2}+\frac{(\frac{4305q^2}{16}+\frac{25875}{128})}{x_0^3}\ee
\be
\Omega_2(x_0)\approx-1+\frac{\frac{3}{8}}{x_0}+\frac{(\frac{1035}{16}+48q^2)}{x_0^2}-\frac{(\frac{4305q^2}{8}+\frac{25875}{64})}{x_0^3}
\ee
\be\Gamma_0(x_0)\approx1+\frac{\frac{15}{8}}{x_0}-\frac{(9q^2+\frac{225}{32})}{x_0^2}+\frac{(\frac{495q^2}{16}+\frac{3375}{128})}{x_0^3}
\ee\be
\Gamma_1(x_0)\approx-\frac{\frac{15}{8}}{x_0}+\frac{(18q^2+\frac{225}{16})}{x_0^2}-\frac{(\frac{1485q^2}{16}+\frac{10125}{128})}{x_0^3}
\ee and
\be\Gamma_2(x_0)\approx-\frac{(9q^2+\frac{225}{32})}{x_0^2}+\frac{(\frac{1485q^2}{16}+\frac{10125}{128})}{x_0^3}.\ee
 Inserting (4.12), (4.13), (4.14), (4.15) and (4.16) into the integral solution
 (4.11) and using some simple calculations, one infers \be I_{eff}^{weak}(x_0>x_{ps}^{eff})
 \approx\pi-\frac{(\frac{33}{8}+\frac{471}{128}+\frac{9\pi q^2}{2})}{x_0}+
 \frac{(\frac{741\pi q^2}{32}+\frac{64863\pi}{2048}-\frac{8841q}{1024}-\frac{997q^2}{16})}{x_0^3}.
 \ee
 Defining \be y=\frac{x_0}{x_{ps}^{eff}}>1\ee and inserting (4.17)
 the deflection angle (4.1) become
 \be \alpha_{eff}^{weak}(y_0>1)\approx-\frac{M}{y_0}+\frac{N}{y_0^3}\ee for weak gravitational lensing where we defined
 \be M(x^{eff}_{ps},q)=\frac{1}{x_{ps}^{eff}}\bigg(\frac{33}{8}+\frac{471}{128}+\frac{9\pi q^2}{2}\bigg),\ee
 and \be N(x_{ps}^{eff},q)=\frac{1}{(x_{ps}^{eff})^3}\bigg(\frac{741\pi q^2}{32}+\frac{64863\pi}{2048}-\frac{8841q}{1024}-\frac{997q^2}{16}\bigg).\ee
 Applying (3.5) and (4.5)  we  obtain  \be
\Omega(x_0)=H,~~~\Omega_1(x_0)=2H-x_0H^{\prime},~~~\Omega_2(x_0)=x_0H^{\prime}-H-x_0^2H^{\prime\prime}/2$$$$
\Gamma(x)=\Gamma_0(x)=1,~~~\Gamma_{1}(x)=0=\Gamma_2(x)\ee
  which are applicable for weak deflection angle in absence of NEM
  field effects. In the latter case asymptotic behavior of the function (4.22) are obtained for $x_0>x_{ps}$  as follows.
  \be \Omega_1(x_0)\approx2-\frac{6}{x_0}+\frac{4q^2}{x_0^2}+\frac{15q^2}{x_0^3}\ee and
  \be\Omega_2(x_0)\approx-1+\frac{6}{x_0}-\frac{6q^2}{x_0^2}-\frac{30q^2}{x_0^3
  }\ee   Inserting  (4.22), (4.23) and (4.24) one can obtain
asymptotic series expansion of the integral solution (4.11) as
follows. \be
I^{weak}(x_0)\approx\pi+\frac{3(\pi-2)}{x_0}+\frac{(8q^2-36-6\pi
q^2+\frac{27\pi}{2})}{x_0^2}+\frac{(123q^2-198-42q^2\pi+\frac{135\pi}{2})}{x_0^3}\ee
which by inserting into (4.1) one can obtain weak deflection angle
of bending light rays in absence of NEM field such that
 \be
\alpha^{weak}(y_0>1)\approx\frac{S}{y_0}+\frac{R}{y_0^2}+\frac{Q}{y_0^3}\ee
where we defined \be
S(x_{ps},q)=\frac{3(\pi-2)}{x_{ps}},~~~R(x_{ps},q)=\frac{(8q^2-36-6\pi
q^2+\frac{27\pi}{2})}{x_{ps}^2}\ee and \be
Q(x_{ps},q)=\frac{(123q^2-198-42q^2\pi+\frac{135\pi}{2})}{x_{ps}^3}.\ee
Diagrams of the equations (4.19) and (4.26) are plotted in figure
2 for ansatz $y_0=10$ by inserting numerical values given in the
table 1. It is suitable to obtain the following averaged
deflection angles.
\be\sigma_{eff}^{weak}=\bar{\alpha}_{eff}^{weak}\approx-\frac{\bar{M}}{y_0}+\frac{\bar{N}}{y_0^3}\ee
and
\be\sigma^{weak}=\bar{\alpha}^{weak}\approx\frac{\bar{S}}{y_0}+\frac{\bar{R}}{y_0^2}+\frac{\bar{Q}}{y_0^3}
\ee where we defined all mean coefficients
\be\bar{\mathcal{C}}=\frac{1}{N}\sum_{i=1}^N\mathcal{C}(x_{ps_i},q_i)\ee
in which $\mathcal{C}=\{M,N,S,R,Q\}.$ Inserting numerical values
of the table 1  we obtain \be
\bar{M}=4.79,~~~\bar{N}=5.41,~~~\bar{S}=1.32,~~~\bar{R}=0.64,~~~\bar{Q}=0.80\ee
for which (4.29) and (4.30) become respectively \be
\sigma_{eff}^{weak}(y_0>1)\approx-\frac{4.79}{y_0}+\frac{5.41}{y_0^3}\ee
and
\be\sigma^{weak}(y_0>1)\approx\frac{1.32}{y_0}+\frac{0.64}{y_0^2}+\frac{0.80}{y_0^3}.\ee
Diagrams of the above mean weak deflection angles are given in the
figure 2. They show that the sign of deflection angle is changed
in presence of NEM fields with respect to sign of deflection angle
in absence of it.
\subsection{Strong lensing
deflection angles}
 In case of strong deflection limits we
write Taylor series expansion of the integral solution (4.11) at
neighborhood of $x_0\to x^{eff}_{ps} (y_0\to1).$ In the latter
case we must be obtain Taylor series expansion of the functions
$\Omega_{1,2}(x_0)$ and $\Gamma_{0,1,2}(x_0)$  which up to terms
in order of $O(3)$ become respectively \be \Omega_1(x_0)\approx
P_{ps}(y_0-1)+Q_{ps}(y_0-1)^2,\ee \be
\Omega_2(x_0)\approx\Omega_2(x_{ps})+R_{ps}(y_0-1)-Q_{ps}(y_0-1)^2\ee
\be
\Gamma_0(x_0)\approx\Gamma_0(x_{ps})+U_{ps}(y_0-1)+V_{ps}(y_0-1)^2,\ee
\be\Gamma_1(x_0)\approx
U_{ps}+(U_{ps}+2V_{ps})(y_0-1)+(2V_{ps}+W_{ps})(y_0-1)^2,\ee and
\be \Gamma_2(x_0)\approx
U_{ps}+V_{ps}+(U_{ps}+4V_{ps}+W_{ps})(y_0-1)+3(V_{ps}+W_{ps})(y_0-1)^2\ee
where we defined \be
P_{ps}=2\Omega_2(x_{ps}),~~~~\Omega_2(x_{ps})=\Omega(x_{ps})-x_{ps}^2\Omega''(x_{ps})/2$$$$
R_{ps}=-x^2_{ps}\Omega^{\prime\prime}(x_{ps}),~~~Q(x_{ps})=-x_{ps}^3\Omega^{'''}(x_{ps})/2
$$$$
U_{ps}=x_{ps}\Gamma_0^{\prime}(x_{ps}),~~~V_{ps}=x_{ps}^2\Gamma_0^{\prime\prime}(x_{ps})/2,~~~
~W_{ps}=x_{ps^3}\Gamma_0^{\prime\prime\prime}(x_{ps})/2.\ee
Inserting (4.35), (4.36), (4.37), (4.38) and (4.39) into the
integral solution
 (4.11) one can obtain strong deflection limits of bending light ray angle (4.1) as follows.
  \be \alpha_{eff}^{strong}(y_0\to1)\approx b+a\ln(y_0-1)\ee where we defined \be b=-\pi+\frac{3U_{ps}+2\ln2
  V_{ps}\Gamma_0(x_{ps})}{\sqrt{\Omega_2(x_{ps})}}\ee
  and \be a=-\frac{2\Gamma_0(x_{ps})}{\sqrt{\Omega_2(x_{ps})}}.\ee Divergency of the above equation
  in limits $y_0=1$ can be described by Bozza formalism as follows.
  \be\alpha_{eff}^{strong}=\alpha_n=\Delta\alpha_n+2n\pi\ee where $n=0,\pm1,\pm2,\pm3,\cdots $
  means $n^{th}$ circulation of light rays around the lens center to make $n^{th}$ relativistic images by deflecting
  $0<\Delta\alpha_n<<1.$ Non-relativistic images are determined by
setting $n=0$ and relativistic images with positive (negative)
parity are determined by setting $n=1,2,\cdots (n=-1,-2,\cdots)$
where one can obtain $\Delta\alpha_{-n}=2\alpha-\Delta\alpha_n$.
In  case of retro-lensing  where observer is located between
source and lens, the light rays come back after than that turning
around the lens (see figure 1 at ref. [40,41]). In the latter case
the parameter $2n$ given in the formula (4.44) must be replaced
with $2n-1.$
    In case of
strong deflection limits in absence of NEM fields we should use
(4.41) but by inserting \be
P_{ps}=2H(x_{ps})-x_{ps}^2H''(x_{ps}),~
~~U_{ps}=-\frac{2x_{ps}H'}{H^3},~~~V_{ps}=-\frac{x_{ps}^2H''}{H^3}+\frac{3x^2_{ps}H'^2}{H^4}
\ee and \be R_{ps}=-x_{ps}^2H'',~~~Q_{ps}=-x_{ps}^3H'''/2.\ee We
now study image locations in weak and strong deflection limits.
\section{Images locations} In order to
calculate the weak deflection images we choose Ohanian lens
equation [47] which has high accuracy and so lower errors with
respect to other lens equations [55]. It has the advantage of
being the closest relative of the exact lens equation, since it
only contains the asymptotic approximation without any additional
assumptions. It can be rewritten  against observational
coordinates as image position $\theta$, source position $\beta$
and deflection angle of bending light rays $\alpha^{weak}$ as
follows (see [55] for more discussions). \be
\arcsin(D_L\sin\theta)-\arcsin(D_S\sin\beta)=\alpha-\theta\ee in
which we defined \be
D_L=\frac{d_{OL}}{d_{LS}},~~~D_S=\frac{d_{OS}}{d_{LS}}.\ee In the
above equations $d_{OS}$ is distance between observer and source,
$d_{OL}$ is distance between the observer and the lens, $d_{LS}$
is distance between the lens and the source. $\theta$ is formed
when a line passing through the observer and the image is coincide
optical axis (line passing through the observer and the lens).
$\beta$ is formed when a line passing through the observer and the
source is coincide the optical axis. One can obtain general
solutions of the lens equation (5.1) as follows.
\be\theta_K(\alpha,\beta)=\arctan\bigg[\frac{D_S\cos\alpha\sin\beta+
K\sin\alpha\sqrt{1-D_S^2\sin^2\beta}}{D_L-D_S\sin\alpha\sin\beta+\cos\alpha\sqrt{1-D_S^2\sin^2\beta}}
\bigg]\ee  where $K=\pm1.$ It has some real solutions for \be
\sin\beta\leq\frac{1}{D_S}.\ee In the following we use (5.3) to
obtain non-relativistic and relativistic image locations.
\subsection{Weak lensing images }
In weak deflection limits with large distances between lens,
source and observer located in a straight line approximately, one
can infer [55] \be D_{S}\approx D_{L}+1.\ee where $D_L>>1$ must be
inserted via experimental date. As a realistic example of
gravitational lens we consider a big black hole located in the
center of Galaxy and study image locations of a star located far
from it. This black hole is called as Sgr A$^*$ [48-51]. Its mass
is estimated as $3.6\times 10^6M_{\bigodot}$ and its distance from
the earth is $d_{OL}=8 kpc=2.47\times 10^{17} m$ with
corresponding Schwarzschild radius $R_{SCH}=10^{10} m.$ We
consider a source to be a star located at distance
$d_{LS}=1.7\times10^{13} m$ from the black hole which is far from
the margin of the accretion disk of the black hole, so it may not
be fall toward
 the black hole center.  For the latter black hole we will have  \be D_L\approx1.45\times10^4\approx D_S.\ee
 The relations (5.4) and (5.6) leads us to choose \be\sin\beta\approx\beta\leq\beta_M,\ee in weak deflection limits
 of gravitational lensing where we defined
 \be\beta_M=\frac{1}{D_S}\approx\frac{1}{D_L}\approx7\times10^{-5}Rad\equiv40107\mu~arc~sec
 \ee  in which the subscript $M$ denotes to the word `Maximum`.
 For critical source $\beta=\beta_M$ the lens equation (5.3) reads \be \theta_M(\alpha)=\arctan\bigg[\frac{\cos\alpha}{D_L-\sin\alpha}
 \bigg]\ee which for weak deflection limits $\alpha\to0$  leads to the following approximation. \be\theta_M\approx\frac{1}{D_L}\approx 40107 \mu~arc~sec.
 \ee  Defining  \be \theta^*=\frac{\theta}{\theta_M},~~~\beta^*=\frac{\beta}{\beta_M}\ee we can obtain
Taylor series expansion of the lens equation (5.3) at neighborhood
of $(|\alpha|,|\beta^*|)<<(1,1)$ as follows. \be
\theta^*_K\approx\frac{KD_L}{1+D_L}\alpha+\frac{D_L}{1+D_L}\beta^*-\frac{K}{6}\frac{D_L^2(D_L-1)
}{(1+D_L)^3}\alpha^3$$$$-\frac{D_L}{2}\frac{D^2_L-2KD_L+2+D_L-2K}{(1+D_L)^3}\beta^*\alpha^2$$$$
-\frac{D_L}{2}\frac{(D_L^2K+D_LK-2D_L+2K-2)}{(1+D_L)^3}\alpha{\beta^*}^2+\frac{D_L}{6}\frac{1+3D_L}{(1+D_L)^3}{\beta^*}^3$$$$
+\frac{KD_L}{120}\frac{(D_L^4+11D_L^2-11D_L^3-D_L)}{(1+D_L)^5}\alpha^5+\cdots\ee
which in limits $D_L\to\infty$ become \be \theta_K^*\approx
K\alpha+\beta^*-\frac{\alpha^2\beta^*}{2}-\frac{K\alpha{\beta^*}^2}{2}-\frac{K}{6}\alpha^3+\frac{K}{120}\alpha^5+\cdots.\ee
 This is primary image location and by transforming $\beta^*\to-\beta^*$ as $\theta_K^*(-\beta^*)$ one can obtain secondary image location.
 The parameter $K=\pm1$ describes right-handed (+1)and/or left-handed (-1
 ) bending angles. Setting $K=+1, \beta^*=1, y_0=10$ and inserting numerical values of the deflection angles (4.19) and (4.26)
 via numerical values of the table 1 (see diagrams of the figure 2) we plot numerical values of $\theta_i^{eff}$ and $\theta_i$
 against different values of $|q_i|$ in figure 3.
 Also we insert (4.33) and (4.34) into the weak lens equation (5.13) by setting $K=+1$ and plot mean weak image locations in figure 3 against $|q|$
  in figure 3.
We now obtain relativistic image locations.
\subsection{Strong lensing images}
The Virbhadra-Ellis lens equation given by [56] \be
\tan\beta=\tan\theta-D[\tan(\alpha-\theta)+\tan\theta]\ee is
useful to study gravitational lensing in strong field limits. In
the above lens equation we have $D=\frac{d_{ls}}{d_{os}}$ for
standard lensing in which lens is located between observer and
source, $D=\frac{d_{os}}{d_{ol}}$ for situations where the source
is located between observer and lens, and with
$D=\frac{d_{os}}{d_{ls}}$ for situations where the observer is
located between the source and the lens (the retro-lensing). The
lensing effects are more important when the objects are highly
aligned, in which $\beta,\theta$ are small and $\alpha$ is close
to $2n\pi+\Delta\alpha_n$ (standard lensing) and/or
$(2n+1)\pi+\Delta\alpha_n$ (retro-lensing) with
$0<\Delta\alpha_n<<1$, $n=1,2,\cdots$. In the latter case we can
use the approximations $\tan\theta\approx\theta$ and
$\tan\beta\approx\beta$ for the lens equation (5.14) and insert
$\alpha=2n\pi+\Delta\alpha_n$ such that (see Eq. 32 in ref. [52]
)\be \beta=\theta- D\Delta\alpha_n.\ee Defining coordinate
independent impact parameter \be u=\frac{r}{\sqrt{\Omega(r)}},\ee
one can obtain its Taylor series expansion as (see Eq.28 in Ref.
[52])\be
y_0-1\approx\sqrt{2}\bigg(\frac{\Omega_2(x_{ps})}{\Omega(x_{ps})}\bigg)^{-\frac{1}{2}}\bigg(\frac{u_0}{u_{ps}}-1\bigg)^\frac{1}{2}\ee
in which $u_0=d_{LO}\sin\theta$ reading to $u_0\approx
d_{LO}\theta$ for small $\theta.$ Inserting the latter relation
and (5.17), the strong deflection angle (4.41) become \be
\alpha(\theta)=c_2-c_1\ln\bigg(\frac{d_{OL}}{u_{ps}}\theta-1\bigg)\ee
where we defined \be
c_1=-\frac{a}{2},~~~c_2=b+\frac{a}{2}\ln\bigg(\frac{2\Omega(x_{ps})}{\Omega_2(x_{ps})}\bigg).\ee
One can obtain $\theta(\alpha)$ by inverting (5.18) as
\be\theta(\alpha)=\frac{u_{ps}}{d_{OL}}(1+e^{(c_2-\alpha)/c_1})\ee
and inserting (4.44) can be rewritten as \be
\theta_n=\frac{u_{ps}}{d_{OL}}(1+e^{(c_2-2n\pi-\Delta\alpha_n)/c_1}).\ee
Making first order Taylor series expansion of the above equation
around $\alpha=2n\pi$ the angular position of $n^{th}$
relativistic image is obtained as
\be\theta_n\approx\theta^{(0)}_n-\zeta_n\Delta\alpha_n\ee where we
defined
\be\theta^{(0)}_n=\frac{u_{ps}}{d_{OL}}[1+e^{(c_2-2n\pi)/c_1}]\ee
and \be\zeta_n=\frac{u_{ps}}{c_1d_{OL}}e^{(c_2-2n\pi)/c_1}.\ee
Eliminating $\Delta\alpha_n$ between (5.15) and (5.22) we obtain
 \be \theta_n=\bigg(1+\frac{\zeta_n}{D}\bigg)^{-1}\bigg(\theta_n^{(0)}+\frac{\zeta_n}{D}\beta\bigg)\ee in which $0<\zeta_n/D<<1$ and so we can
 use approximation $(1+\frac{\zeta_n}{D})^{-1}\approx1-\frac{\zeta_n}{D}.$ In the latter case the equation (5.25) reads
 \be\theta_n\approx\theta_n^{(0)}+\frac{\zeta_n}{D}(\beta-\theta_n^{(0)}).\ee The second term in the
 above lens equation is more smaller than the first term which means all relativistic image locations lie very close to $\theta_n^{(0)}.$
 There are other set of relativistic images by changing
 $\theta_n^{(0)}\to-\theta_n^{(0)}$ into the above lens equation.
 In case of perfect aligned $\beta=0$ the above lens equation
 reaches to \be \theta_n^E=\bigg(1-\frac{\zeta_n}{D}\bigg)\theta_n^{(0)}\ee
 describing $n^{th}$ relativistic Einstein ring.
  \subsubsection{Magnifications} The
magnification $\mu$ of an image is defined as the ratio of flux of
the image to flux of un-lensed source. It has two components
called as tangential $\mu_t=\frac{\sin\theta}{\sin\beta}$ and
radial $\mu_r=\frac{d\theta}{d\beta}$ which their multiplication
makes the magnification as
\begin{equation}
\mu=\bigg|\frac{\sin\beta}{\sin\theta}\frac{d\beta}{d\theta}\bigg|^{-1}.
\end{equation}
The above equation denotes to primary image $\theta^p(\beta)$
magnifications with positive parity. Inserting the secondary image
location $\theta^{s}(\beta)= \theta^{p}(-\beta)$ into the
magnification equation (5.28) one can obtain secondary image
magnification  $\mu^{s}_{weak}(\beta)=\mu^{p}_{weak}(-\beta)$ with
negative parity. In the micro-lensing state two weak field images
are not resolved and so the main observables
 should be considered become the total magnification $\mu_{tot}$ and magnification-
weighted-centroid $\mu_{cent}$ defined by respectively [57]
\be\mu_{tot}=|\mu_{s}|+|\mu_{p}|\ee and \be
\mu_{cent}=\frac{\theta_{p}|\mu_{p}|+
\theta_{s}|\mu_{s}|}{|\mu_{p}|+ |\mu_{s}|}.\ee We now calculate
the above magnifications for weak and strong lensing.
\subsection{Weak lensing
magnifications} In case of weak deflection limits we can use
$\sin\theta\approx\theta$ and $\sin\beta\approx\beta$ to evaluate
(5.28) as follows.
\begin{equation}
\mu^{weak}\approx\bigg|\frac{\beta}{\theta}\frac{d\beta}{d\theta}\bigg|^{-1}
\end{equation}
 which by inserting (5.13) reads  \be
\mu_K^{weak}\approx\bigg|K\alpha+\beta^*-
\frac{\alpha^2\beta^*}{2}-\frac{K\alpha\beta^{*2}}{2}-\frac{K\alpha^3}{6}+\frac{K\alpha^5}{120}\bigg|\times\bigg|\frac{1}{
\beta^*}-\frac{\alpha^2}{2\beta^*}-K\alpha\bigg|.\ee Inserting
numerical values given in the table 2 we calculate numerical
values of $\mu, \mu_{tot}, \mu_{cent}$ to plot their diagrams
against $\beta^*$ in figure 4.  Diagrams show that $\mu$ decreases
by increasing $|q|$ in presence and absence of NEM fields.
$\mu_{tot}$ has minimum value for $\beta>0 (\beta<0)$ in presence
(absence) of NEM fields. While corresponding $\mu_{cent}$ take
maximum value. Furthermore we see from the figure 4 that
magnification of the Einstein rings is major in presence of NEM
fields with respect to situations where there is not. We now study
strong lensing magnifications.
\subsubsection{Strong lensing magnifications}
 We now are in position to calculate
$n^{th}$ relativistic images magnification which is determined by
 \be
\mu_n\approx\bigg|\frac{\beta}{\theta_n}\frac{d\beta}{d\theta_n}\bigg|^{-1}\ee
and by inserting (5.26) reads \be
\mu_n\approx\frac{\zeta_n}{D}\frac{\theta_n^{(0)}}{\beta}.\ee The
above magnification is valid for a point source and for  extended
source there is obtained a different form of the magnification
 (see for instance Eq. (46) in Ref. [52]). The equation (5.34) shows that the first relativistic image
 is brightest one, and the magnifications decreases exponentially with `n`.
 Magnifications in case of retro-lensing is obtained by (5.34) but by changing $2n\to2n-1.$
The total magnification, taking into account both sets of
relativistic images, is defined by
$\mu_{tot}^{strong}=2\sum_{n=1}^{\infty}\mu_n.$ Using the formula
$\frac{t}{1-t}=\sum_{n=1}^{\infty}t^n$ and inserting (5.23),
(5.24) and (5.34) we obtain \be
\mu_{tot}^{strong}=\frac{1}{\beta}\bigg(\frac{2u_{ps}^2e^{\frac{c_2}{c_1}}}{Dc_1d^2_{OL}}
\bigg)\bigg[\frac{1+e^{\frac{c_2}{c_1}}+e^{\frac{2\pi}{c_1}}}{e^{\frac{4\pi}{c_1}}-1}\bigg]\ee
for a point source where \be
u_{ps}(x_{ps})=\frac{x_{ps}}{\sqrt{\Omega(x_{ps})}},\ee \be
c_1(x_{ps})=\frac{\Gamma_0(x_{ps})}{\sqrt{\Omega_2(x_{ps})}},\ee
and \be
c_2(x_{ps})=-\pi+\frac{3U_{ps}+V_{ps}+2\ln2\Gamma_0(x_{ps})}{\sqrt{\Omega_2(x_{ps})}}-\frac{\Gamma_0(x_{ps})}{\sqrt{\Omega_2(x_{ps})}}
\ln\bigg[\frac{2\Omega(x_{ps})}{\Omega_{2}(x_{ps})}\bigg]. \ee The
equations (5.37) and (5.38) are obtained by inserting
(4.42) and (4.43) into the relations (5.19).\\
As an example we obtain the lensing observable defined by Bozza
[37] as \be s=\theta_1-\theta_{\infty}\ee and \be
r=\frac{\mu_1}{\sum_{n=2}^{\infty}\mu_n}\ee which are useful when
the outermost relativistic image can be resolved from the rest.
$`s`$ represents the angular separation between the first image
and the limiting value of the succession of images. $`r`$ is the
ratio between the flux of the first image and sum of the fluxes of
the other images. Applying (5.24) and (5.26) one can obtain exact
form of the equation (5.39) which up to term of
$0<\frac{\zeta_1}{D}<<1$ become \be
s\approx\theta_{\infty}e^{(c_2-2\pi)/c_1}\ee in which
\be\theta_{\infty}=\frac{u_{ps}}{\tilde{d}_{OL}}\ee and numerical
value of  $\tilde{d}_{LO}$ is used here for Sgr $A^*$ observed
galactic  black hole as \be
\tilde{d}_{OL}=\frac{d_{OL}}{R_{SCH}}=\frac{2.47\times10^{17}m}{10^{10}m}=2.47\times10^7\ee
One infers that the equation (5.40) can be rewritten as \be
r=\frac{2\mu_1}{\mu_{tot}^{strong}-2\mu_1}\ee which by inserting
(5.34) and (5.35) reads \be
r=\frac{e^{\frac{2\pi}{c_1}}+e^{\frac{c_2}{c_1}}-e^{-\frac{2\pi}{c_1}}+e^{\frac{c_2}{c_2}}e^{-\frac{4\pi}{c_1}}}{1+e^{-\frac{2\pi}{c_1}}
-e^{\frac{c_2}{c_1}}e^{-\frac{4\pi}{c_1}}} \ee and its Taylor
series expansion about $e^{-\frac{2\pi}{c_1}}\to0$ become \be
r\approx e^{\frac{2\pi}{c_1}}+e^\frac{c_2}{c_1}-1.\ee We plot
diagrams of $c_{1,2}$ against $|q|$ in figure 5 for $|q|<1$ and
figure 7 for $|q|>1.$ They show raise of $c_{1,2}$ to  large
negative values in absence of NEM fields. While in presence of NEM
fields $c_1$ takes some positive values for $|q|<0.6$ and reaches
to negative values for $0.6<|q|<1.$ $c_2$ decreases by increasing
$|q|<1$ in presence of NEM fields. For $|q|>1$ the diagrams of the
figure 7 shows  increase of $c_1 (c_2)$ to some positive
(negative) values by decreasing $|q|>1$ in absence (presence) of
NEM fields.  $r (s)$ behaves as increasing (decreasing) function
by raising $|q|>1.$ In figure 6 we see increase (decrease) of $s$
by raising $|q|<1$ in absence (presence) of NEM fields.  $r$
decreases (increase) by raising $|q|<1$  in absence (presence) of
the NEM fields.
\section{Concluding remark}
As a black hole solution of Born Infeld Einstein-non linear
Maxwell gravity we use ABG nonsingular charged black hole to study
its gravitational lensing in weak and strong deflection limits. We
set our calculations to Sgr A$^*$ observed black hole date and
study non linear counterpart of EM field on the gravitational
lensing. In short we obtained negligibility of NEM effects for
large values of the black hole charge but not for its small
values. Nonlinearity causes to be larger the photon sphere radius.
Sign of deflection angle of bending light rays is changed.
Einstein rings become larger and their magnifications become
greater. Angular separation of innermost and outermost
relativistic images decreases by increasing the charge parameter
but their magnifications increase. As a future work one can use
strategy and results of this work to compare with results obtained
by studying gravitational lensing of other nonsingular galactic
black holes described in the introduction section and setting
other observed black holes data.

 \vskip0.5 cm {\bf
References}
\begin{description}
  \item[1.]S. W. Hawking and G. F. R. Ellis,
  \emph{The large scale structure of space-time}, (Cambridge University Press, England, 1973).
\item[2.] R. M. Wald, \textit{General relativity}, (The University
of Chicago, Chicago Press, 1984).
\item[3.] R. M. Wald,
 \textit{ Gravitational collapse and cosmic censorship}, In black holes, gravitational radiation and the Universe,
  (Springer, Netherlands, 1999).
\item[4.] I. Dymnikova, ``Spherically symmetric space-time with the regular de Sitter center,''
  Int.  J.  Mod. Phys. D{\bf12}, 1015 (2003).
 \item[5.] A. A. Tseytlin, ``On singularities of spherically symmetric backgrounds in string theory,''
  Phys. Lett.  B{\bf363}, 223 (1995).
\item[6.] M. Cvetic,  ``Flat world of dilatonic domain walls,''
  Phys.  Rev. Lett. {\bf 71}, 815 (1993).
\item[7.] J. H. Horne and G. T. Horowitz,  ``Exact black string solutions in three-dimensions,''
  Nucl. Phys. B{\bf 368}, 444 (1992).
\item[8.] K. A. Bronnikov, V. N. Melnikov and H. Dehnen,  ``Regular black holes and black universes,''
  Gen.  Rel.  Grav.  {\bf 39}, 973 (2007).
\item[9.] S. Ansoldi,  ``Spherical black holes with regular center``, gr-qc/0802.0330 (2008).
\item[10.] J. Bardeen, Proceedings of GR5, Tiflis, USSR, (1968).
\item [11.] E. Ayon-Beato and A. Garcia,  ``The Bardeen model as a nonlinear magnetic monopole,''
  Phys.  Lett.  B{\bf 493}, 149 (2000); gr-qc/0009077.
 \item[12.] E. Ayon-Beato and A. Garcia,  ``Regular black hole in general relativity coupled to nonlinear electrodynamics,''
  Phys.  Rev.  Lett. {\bf 80}, 5056 (1998), gr-qc/9911046v1.
 \item [13.] E. Ayon-Beato and A. Garcia,  ``New regular black hole solution from nonlinear electrodynamics,''
  Phys.  Lett.  B{\bf 464}, 25 (1999).
  \item[14.] E. Ayon-Beato and A. Garcia, ``Nonsingular charged black hole solution for nonlinear source,''
  Gen.  Rel.  Grav. {\bf 31}, 629 (1999).
\item[15.] S. A. Hayward,  ``Formation and evaporation of regular black holes,''
  Phys.  Rev.  Lett. {\bf 96}, 031103 (2006).
\item[16.] J. C. S. Neves and A. Saa,  ``Regular rotating black holes and the weak energy condition,''
  Phys.  Lett.  B{\bf 734}, 44 (2014); gr-qc/1402.2694.
\item[17.] J. C. S. Neves, ``Note on regular black holes in a brane world,''
    Phys.  Rev.  D{\bf 92}, 084015 (2015); gr-qc/1508.0361.
\item[18.] M. Azreg-Ainou,  ``Generating rotating regular black hole solutions without complexification,''
  Phys.  Rev.  D{\bf 90}, 064041 (2014).
\item[19.] C. Bambi and L. Modesto,  ``Rotating regular black holes,''
  Phys.  Lett.  B{\bf 721}, 329 (2013); gr-qc/1302.6075.
\item[20.] B. Toshmatov, B. Ahmedov, A. Abdujabbarov and
Z. Stuchlik,  ``Rotating regular black hole solution,''
  Phys.  Rev.  D{\bf 89}, 104017 (2014); gr-qc/1404.6443.
\item[21.] Sushant G. Ghosh  ``A nonsingular rotating black hole,''
EPJ. C{\bf{75}}, 532, (2015).
\item[22.] E. F. Eiroa and C. M. Sendra,  ``Regular phantom black hole gravitational lensing,''
  Phys.  Rev.  D{\bf 88}, 103007 (2013).
\item[23.] L. Modesto and P. Nicolini,  ``Charged rotating noncommutative black holes,''
  Phys.  Rev.  D{\bf 82}, 104035 (2010); gr-qc/1005.5605.
\item[24.] Sushant G. Ghosh, P. Sheoran and M. Amir  ``Rotating Ayon-Beato-Garcia black hole as a particle accelerator,''
  Phys.  Rev.  D{\bf 90},103006 (2014); gr-qc/1410.5588.
  \item[25.] G. V. Kraniotis, ``Gravitational lensing and frame dragging of light in the Kerr-Newman and the Kerr-Newman-(anti)
   de Sitter black hole space time``, Gen. Rel. Grav. 46 {\bf 11}, 1818 (2014).
\item[26.] M. Patil and P. S. Joshi,  ``Ultra-high energy particle collisions in a regular space time without black holes or naked singularities,''
  Phys.  Rev.  D{\bf 86}, 044040 (2012).
\item[27.] Z. Stuchlik, J. Schee and A. Abdujabbarov,  ``Ultra-high-energy collisions of particles in the field of near-extreme Kehagias-Sfetsos naked singularities and their appearance to distant observers,''
  Phys.  Rev.  D{\bf 89}, 104048 (2014).
\item[28.]A. García, E. Hackmann, J. Kunz, C. Lämmerzahl and A. Macías,  ``Motion of test particles in a regular black hole space–time,''
  J.  Math.  Phys. {\bf 56}, 032501 (2015).
\item[29.] Z. Stuchlik and J. Schee,``Circular geodesic of Bardeen and Ayon-Beato-Garcia regular
black-hole and no-horizon spacetimes,'' Int. J. Mod. Phys. D{\bf
24}, 1550020 (2015).
\item[30.] Z. Stuchlik and J. Schee,  ``Optical effects related to Keplerian discs orbiting Kehagias-Sfetsos naked singularities,''
  Class.  Quant.  Grav. {\bf 31}, 195013 (2014).
\item[31.] J. Schee and Z. Stuchlik, ``Profiled spectral lines generated by Keplerian discs orbiting in the Bardeen and Ayon
 -Beato-Garcia space times``, Class. Quantum Grav.,
   {\bf 33}, 085004,  (2016).
\item[32.] A. Kehagias and K. Sfetsos,  ``The Black hole and FRW geometries of non-relativistic gravity,''
  Phys.  Lett.  B{\bf 678}, 123 (2009).
\item[33.] E. F. Eiroa, G. E. Romero and D. F. Torres,  ``Reissner-Nordstrom black hole lensing,''
  Phys.  Rev.  D{\bf 66}, 024010 (2002).
\item[34.] E. F. Eiroa and C. M. Sendra,  ``Gravitational lensing by a regular black hole,''
  Class.  Quant.  Grav.  {\bf28}, 085008 (2011).
\item[35.] S. W. Wei, Y. X. Liu and C. E. Fu,  ``Null geodesics and gravitational lensing in a nonsingular spacetime,''
  Adv.  High Energy Phys. {\bf 2015}, 454217 (2015).
\item [36.] H. Ghaffarnejad and H. Niad,  ``Weak gravitational lensing from regular Bardeen black holes,''
  Int.  J.  Theor.  Phys. {\bf 54}, 9, 1 (2015); gr-qc/1411.7247.
\item[37.] V. Bozza, ``Gravitational lensing in the strong field limit,''
  Phys.  Rev. D{\bf 66}, 103001 (2002); gr-qc/0208075.
  \item[38]  V. Bozza,  ``Gravitational lensing by black holes,''
  Gen.  Rel. Grav.  {\bf 42}, 2269 (2010).
\item[39.]
  S. Sahu, K. Lochan and  D. Narasimha, `Gravitational lensing by self-dual black hole in loop quantum
  gravity` Phys. Rev. D{\bf 91}, 063001 (2015).
\item[40.] E. F. Eiroa and D. F. Torres,  ``Strong field limit analysis of gravitational retro lensing,''
  Phys.  Rev.  D{\bf 69}, 063004 (2004); gr-qc/0311013.
  \item[41.] E. F. Eiroa, `Braneworld black hole gravitational lensing: Strong field limits
  analysis`, Phys. Rev D{\bf71}, 083010 (2005).
\item[42.] S. Sahu, M. Patil, D. Narasimha and P. S. Joshi,  ``Can strong gravitational lensing distinguish naked singularities from black holes?,''
  Phys.  Rev. D{\bf 86}, 063010 (2012).
\item [43.] R.R.Cuzinato, C.A.M.de Melo, K.C. de Vasconcelos,
L. G. Medeiros and P. J. Pompeia, `` Nonlinear effects on
radiation propagation around a charged compact object``,
Astrophys. Space Sci 359, 59 (2015).
 \item [44.] M. Novello, V. A.
de Lorenci, J. M. Salim and R. Klippert, `` Geometrical aspects of
light propagation in nonlinear electrodynamics``, Phys. Rev. D61,
045001 (2000).
\item [45.] N. Breton, `` Geodesic structure of the Born-Infeld
black hole`` Class. Quantum Grav. 19, 601, (2002).
\item [46.] H. Ghaffarnejad  ``Classical and quantum Reissner Nordstr\"{o}m black hole thermodynamics and first order phase transition,''
  Astrophys. Space Sci. {\bf 361}, 7, 1 (2016); physics.gen-ph/1308.1323.
  \item[47.] H. C. Ohanian, ``The black holes as a gravitational
  lens``, Am. J. Physics {\bf 55} (5), 428 (1987).
\item[48.] R. Genzel et al, ``The
Galactic Center massive black hole and nuclear star cluster``,
Rev.Mod. Phys. {\bf 82} 3121-95 (2010).
\item[49.] A. M. Ghez et al, ``Measuring
distance and properties of the Milky Way's central supermassive
black hole with stellar orbits``, Astrophys. J. {\bf 689}, 1044
(2008).
\item[50.] F. Melia \textit{ The black hole at the center of our
Galaxy}, (Princeton University Press, Princeton 2003).
\item[51.] F. Eisenhauer et al., `` SINFONI in the Galactic Center: Young Stars and Infrared Flares in the Central Light-Month``
,  Astrophys. J. {\bf 628}, 246 (2005).
\item [52.] E. F. Eiroa, `` Gravitational lensing by
Einstein-Born-Infeld black holes`` Phys. Rev. D73, 043002 (2006);
gr-qc/0511065v2.
\item[53.] C. M. Claudel, K. S. Virbhadra and G. F. R. Ellis,  ``The geometry of photon surfaces,''
  J.  Math. Phys. {\bf 42}, 818 (2001).
\item[54.]  S. Weinberg, \textit{Gravitation and
Cosmology: Principle and Applications of the General Theory of
Relativity (Wiley New York 1972).}
\item [55.] V. Bozza, `` Comparison of approximate gravitational
lens equations and a proposal for an improved new one`` Phys. Rev.
D78, 103005 (2008).
\item[56.] K. S. Virbhadra and G. F. R.
Ellis,``Schwarzschild black hole lensing``, Phys. Rev. D62, 0840
03 (2000).
\item [57] B. S. Gaudi and A. O. Petters, `` Gravitational Microlensing Near Caustics II: Cusps``, Astrophys. J. 580, 468-489 (2002),
astro-ph/0206162.
\end{description}

\begin{figure} \centering
\includegraphics[width=2.5in,height=2.0in]{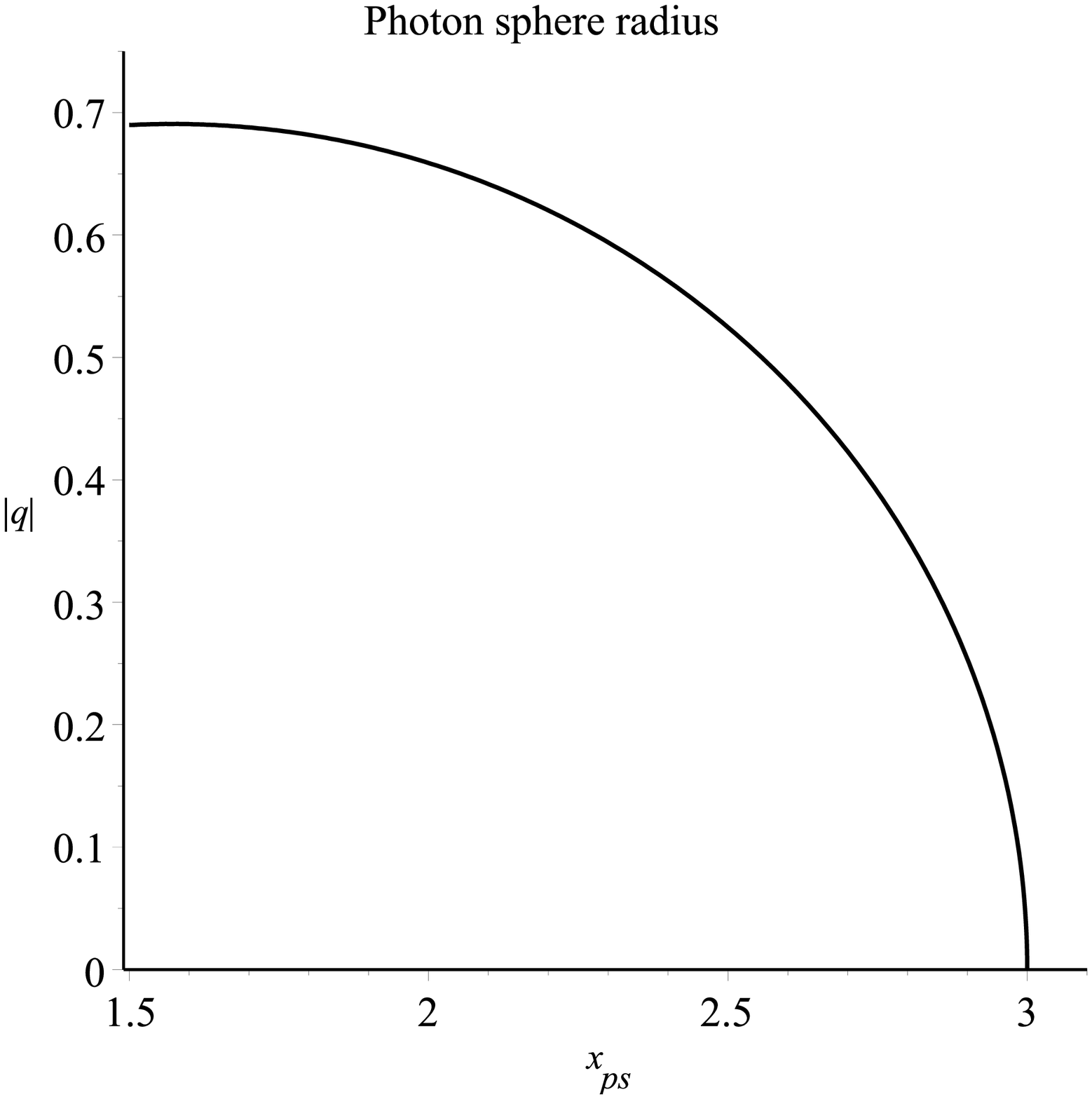}
\includegraphics[width=2.5in,height=2.0in]{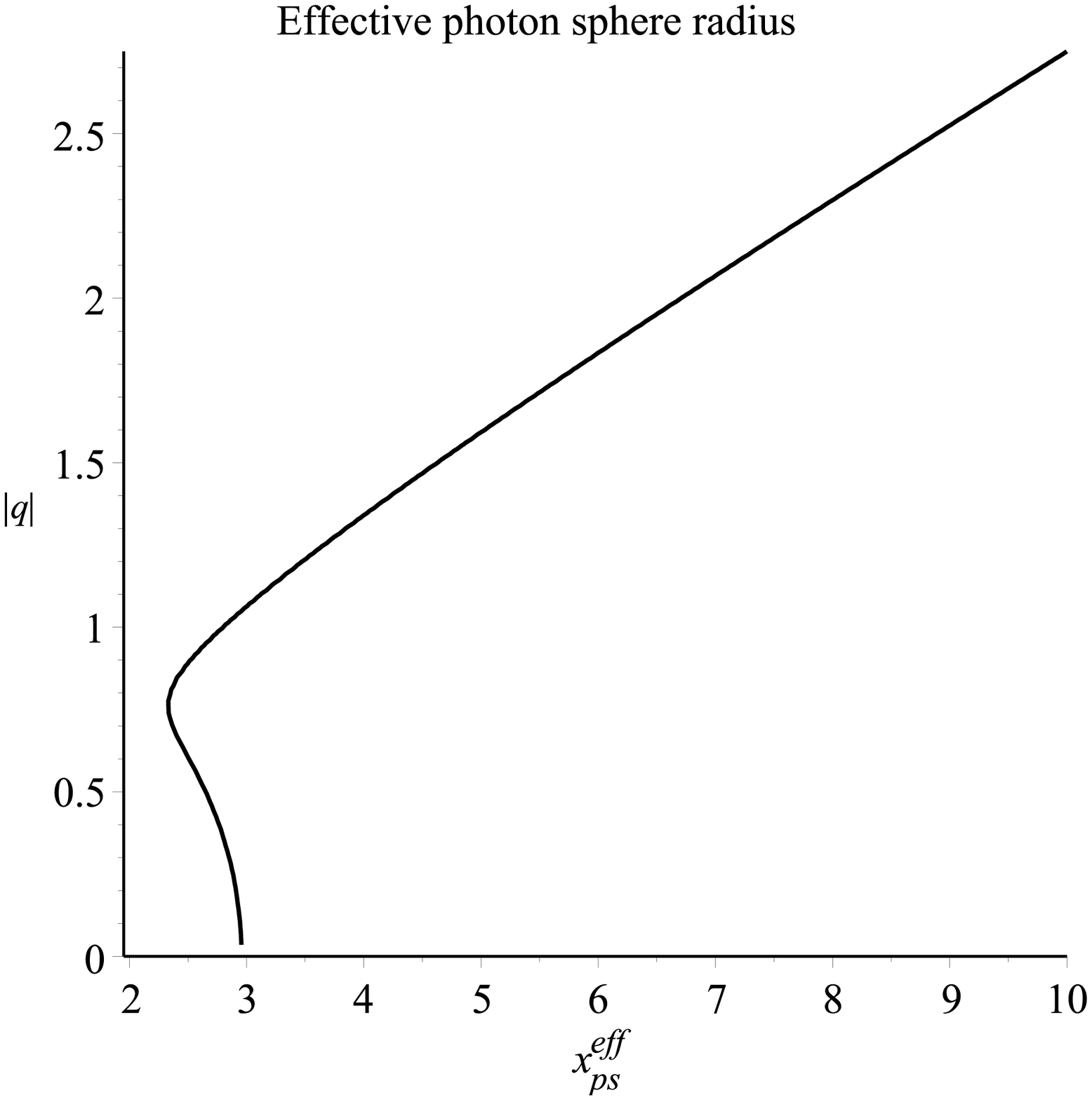}
\includegraphics[width=2.5in,height=2.0in]{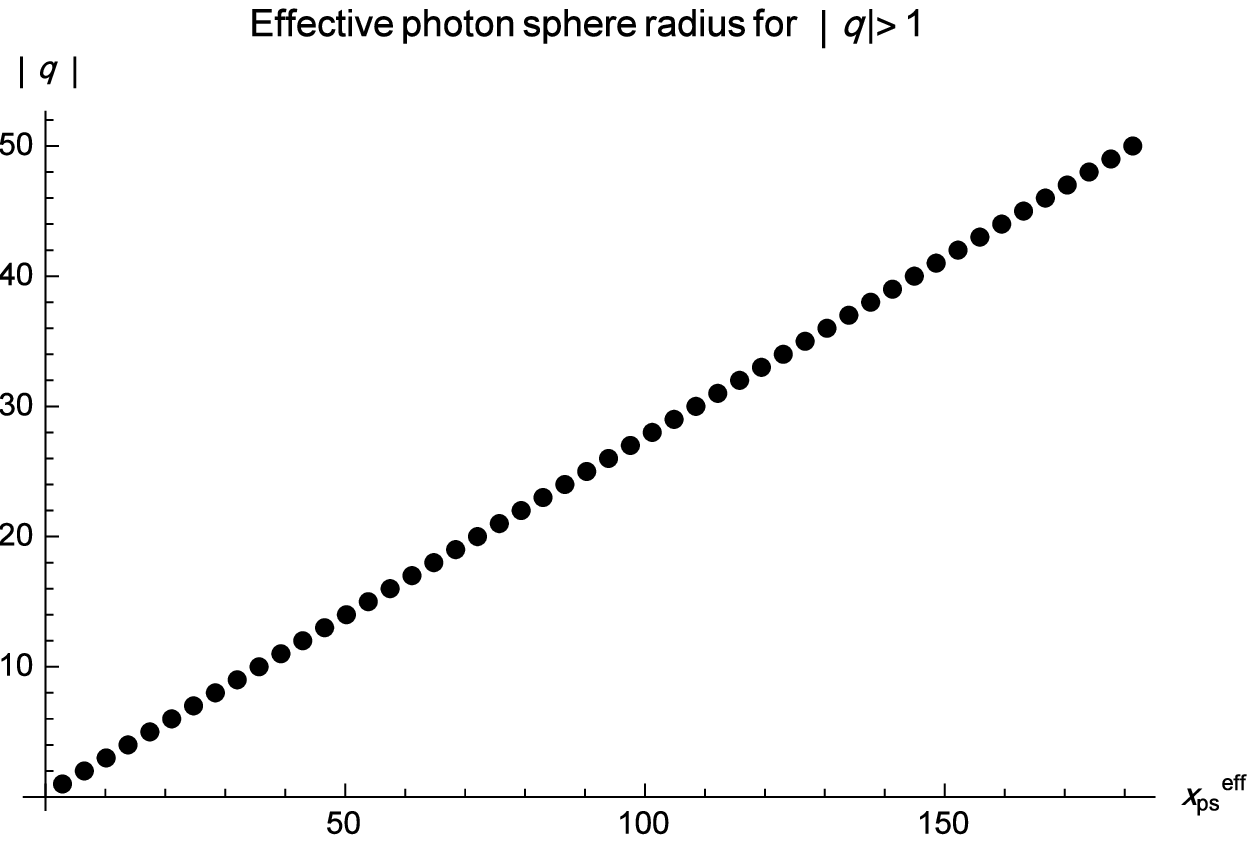}
\includegraphics[width=2.5in,height=2.0in]{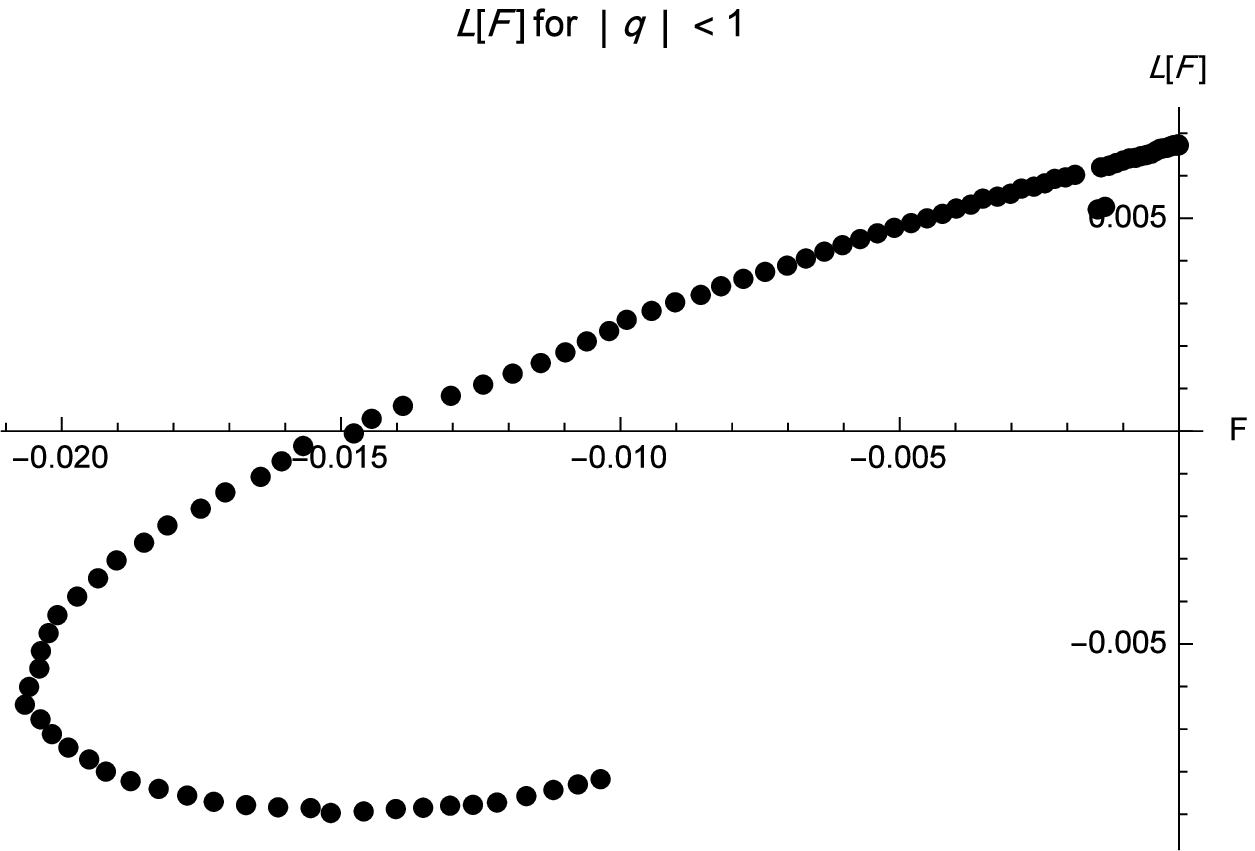}
\includegraphics[width=2.5in,height=2.0in]{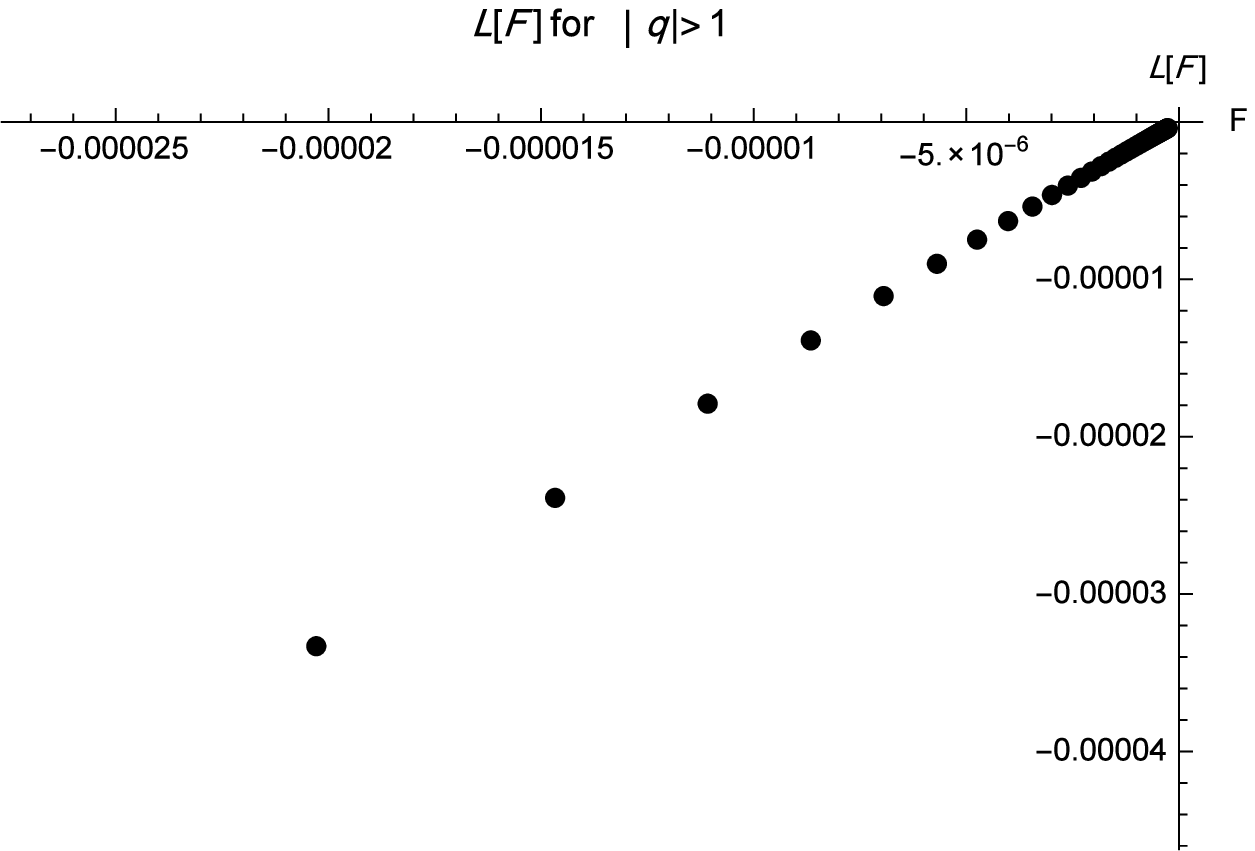}

\caption{{\small \ Diagrams of photon sphere locations $x_{ps}$
and NEM field lagrangian density $L[F]$ }}
\end{figure}

\begin{figure} \centering
\includegraphics[width=2.5in,height=2.0in]{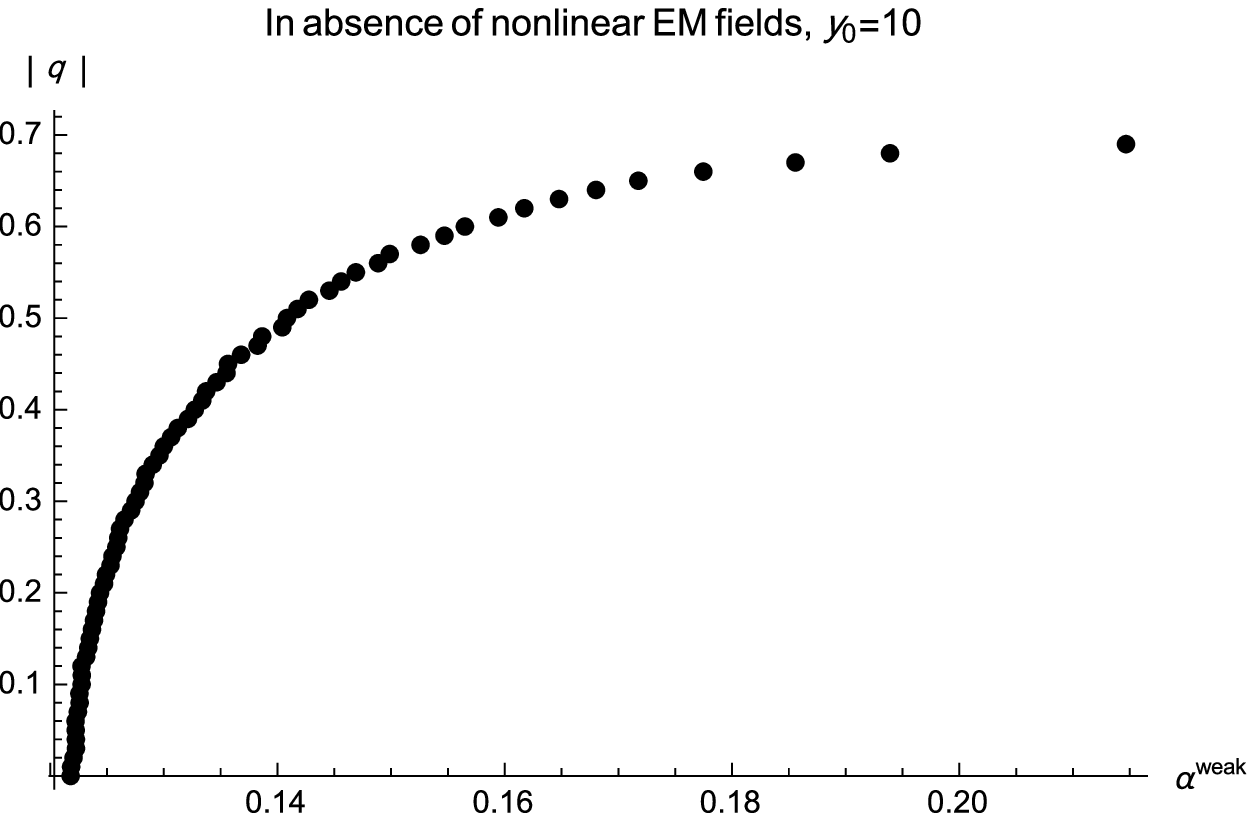}
\includegraphics[width=2.5in,height=2.0in]{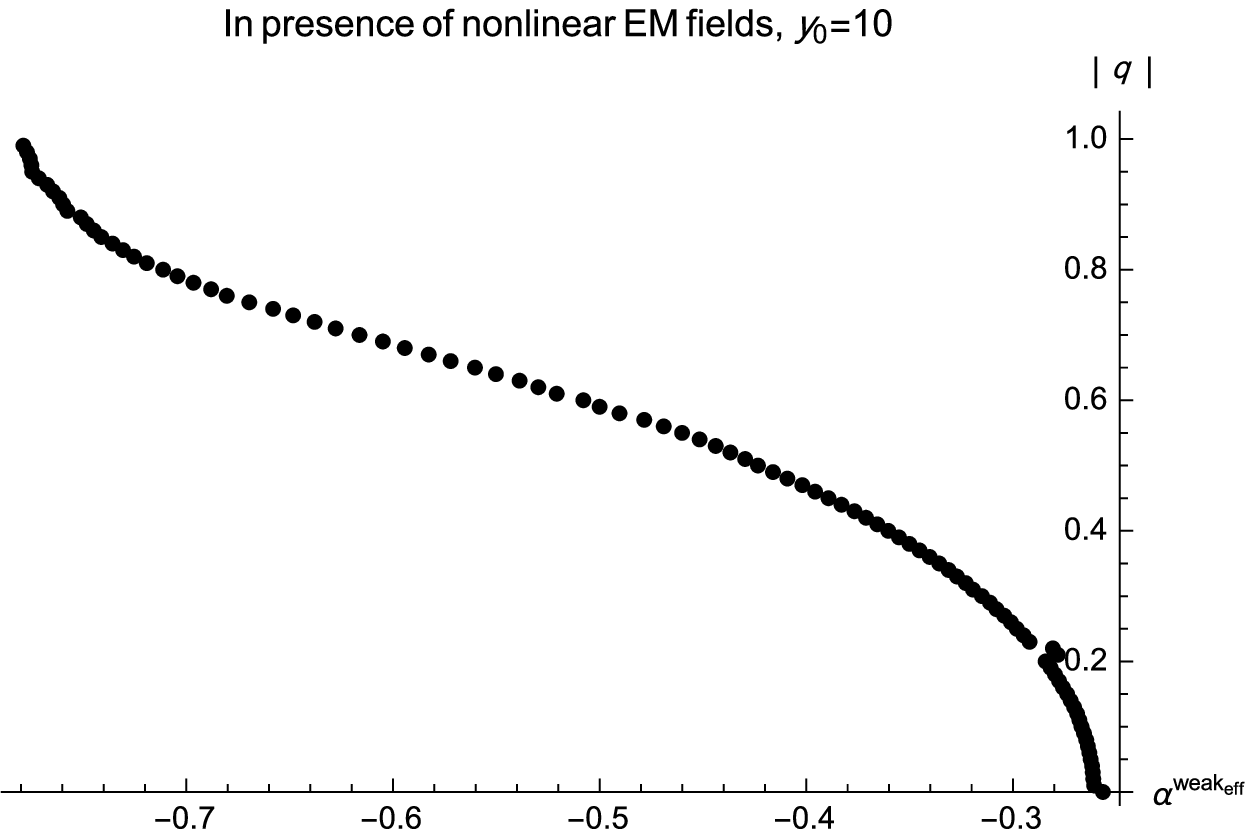}
\includegraphics[width=2.5in,height=2.0in]{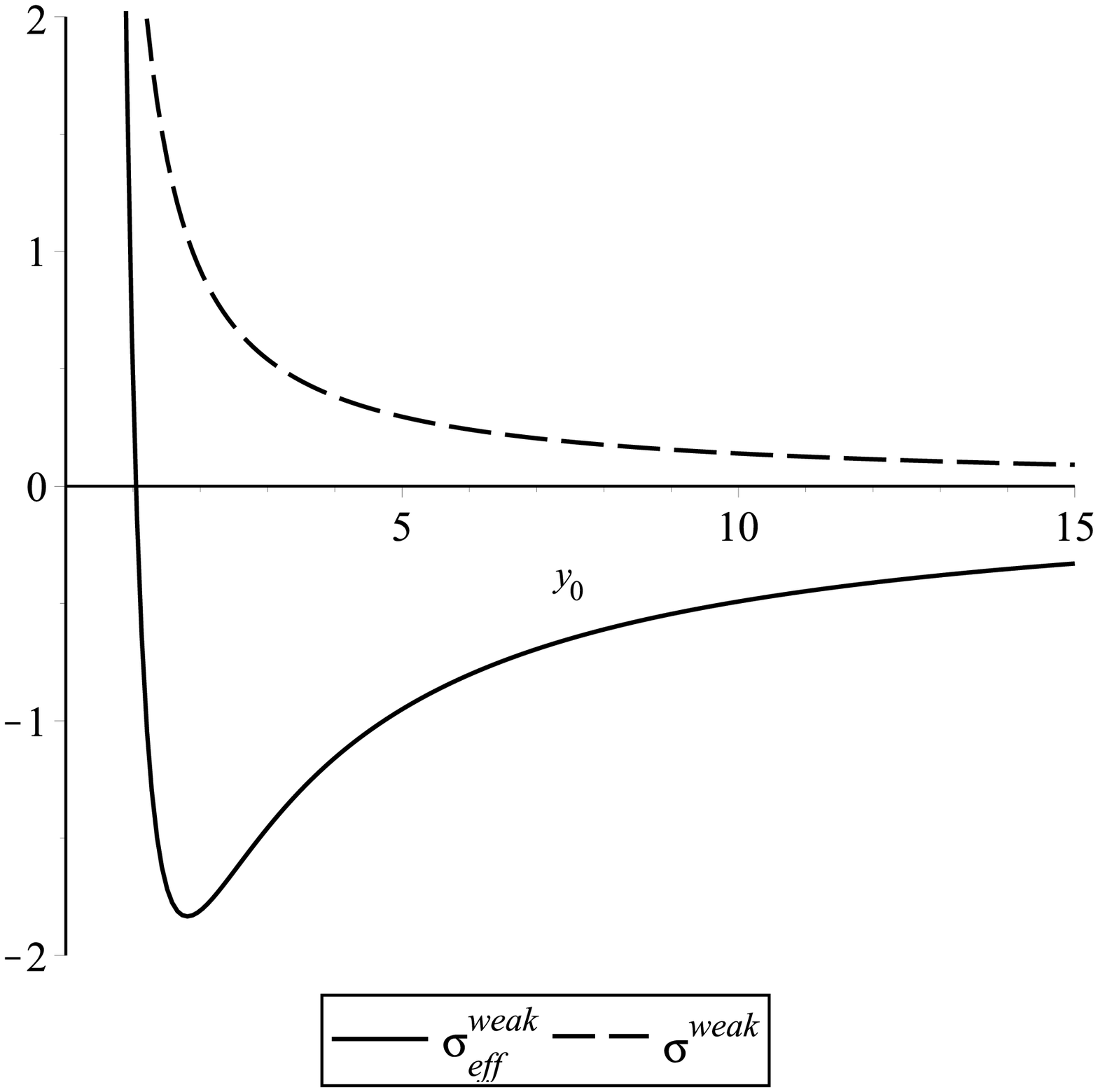}
\includegraphics[width=2.5in,height=2.0in]{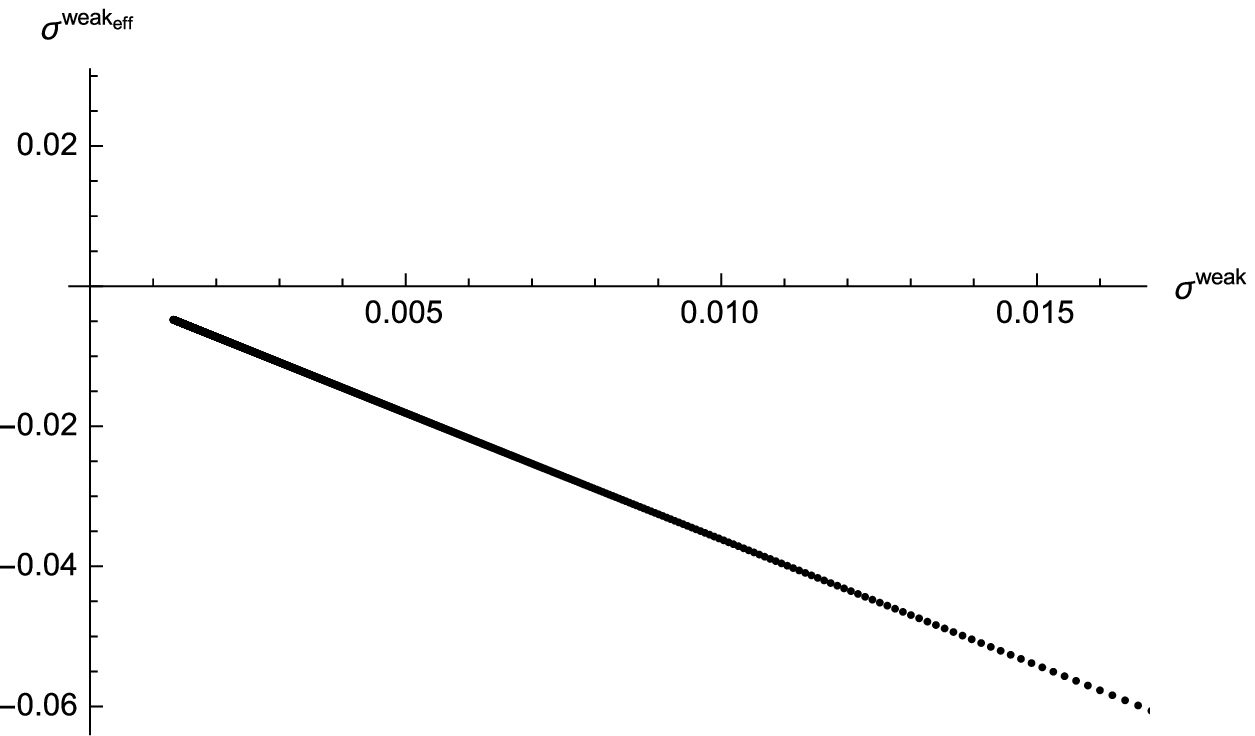}
\caption{{\small \ Diagrams of weak lensing deflection angles for
$|q|<1$ }}
\end{figure}

\begin{figure} \centering
\includegraphics[width=2.5in,height=2.5in]{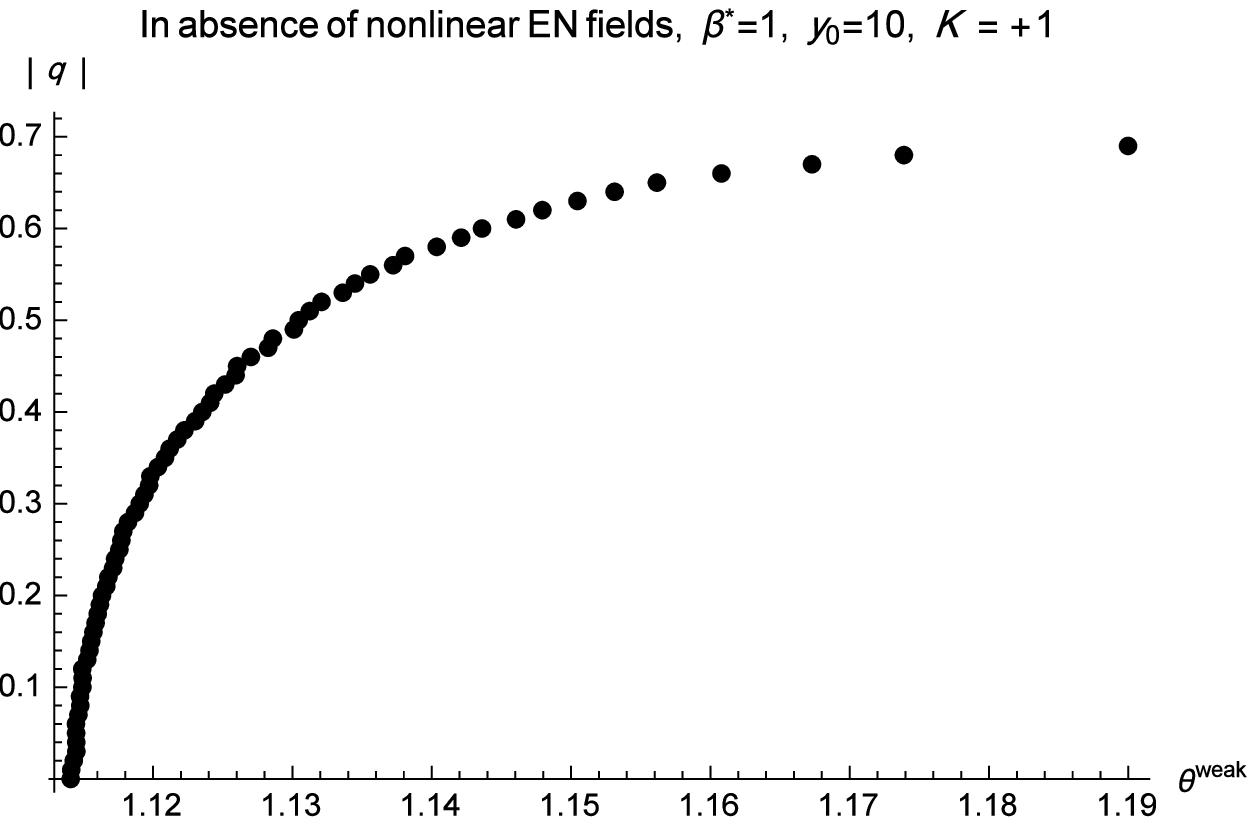}
\includegraphics[width=2.5in,height=2.5in]{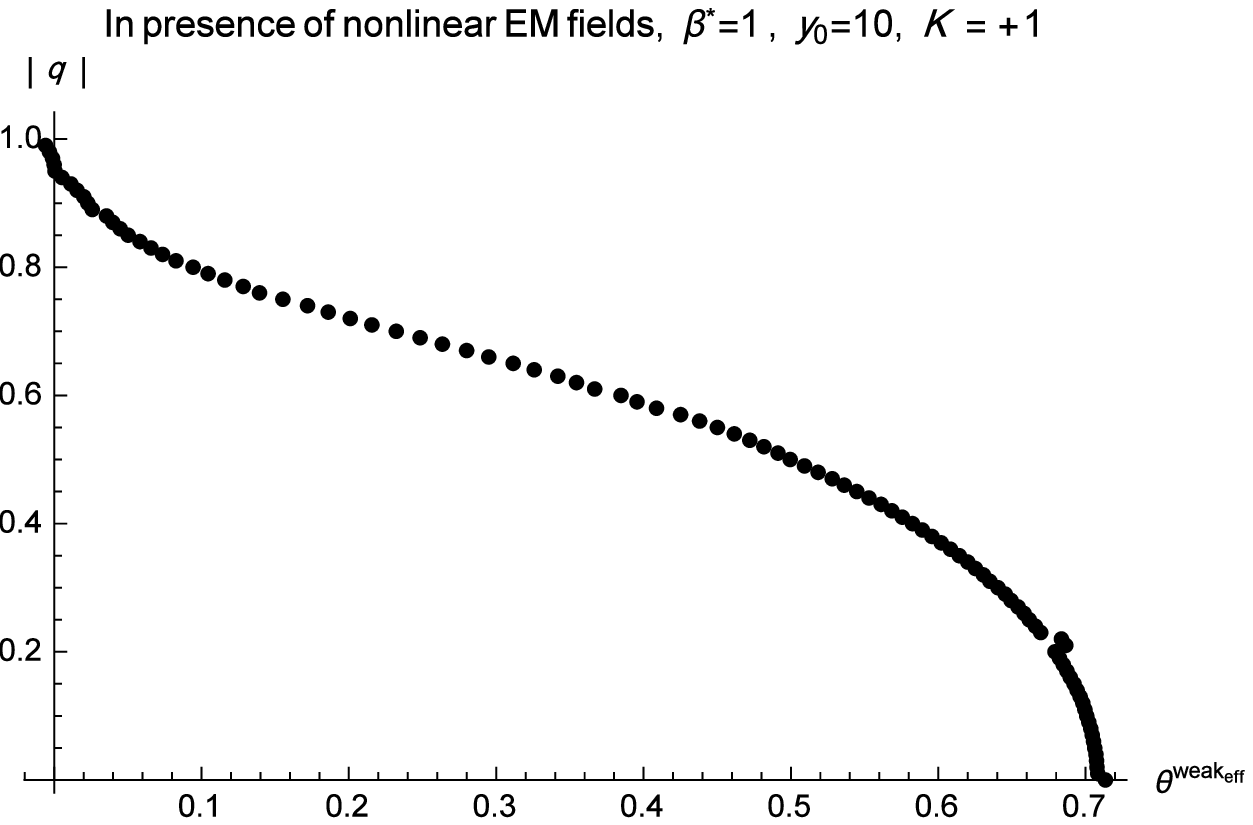}
\includegraphics[width=2.5in,height=3in]{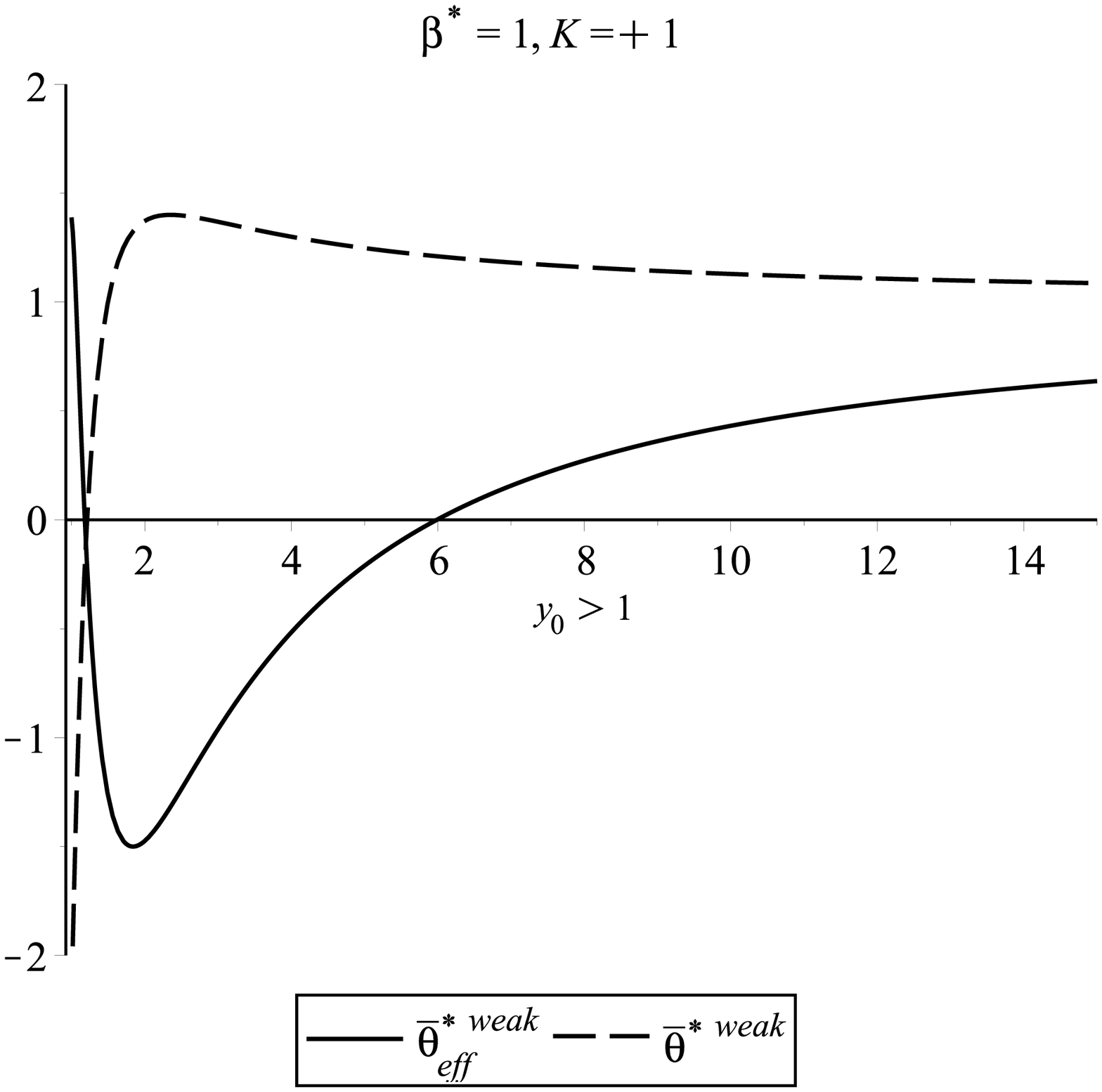}
\includegraphics[width=2.5in,height=3in]{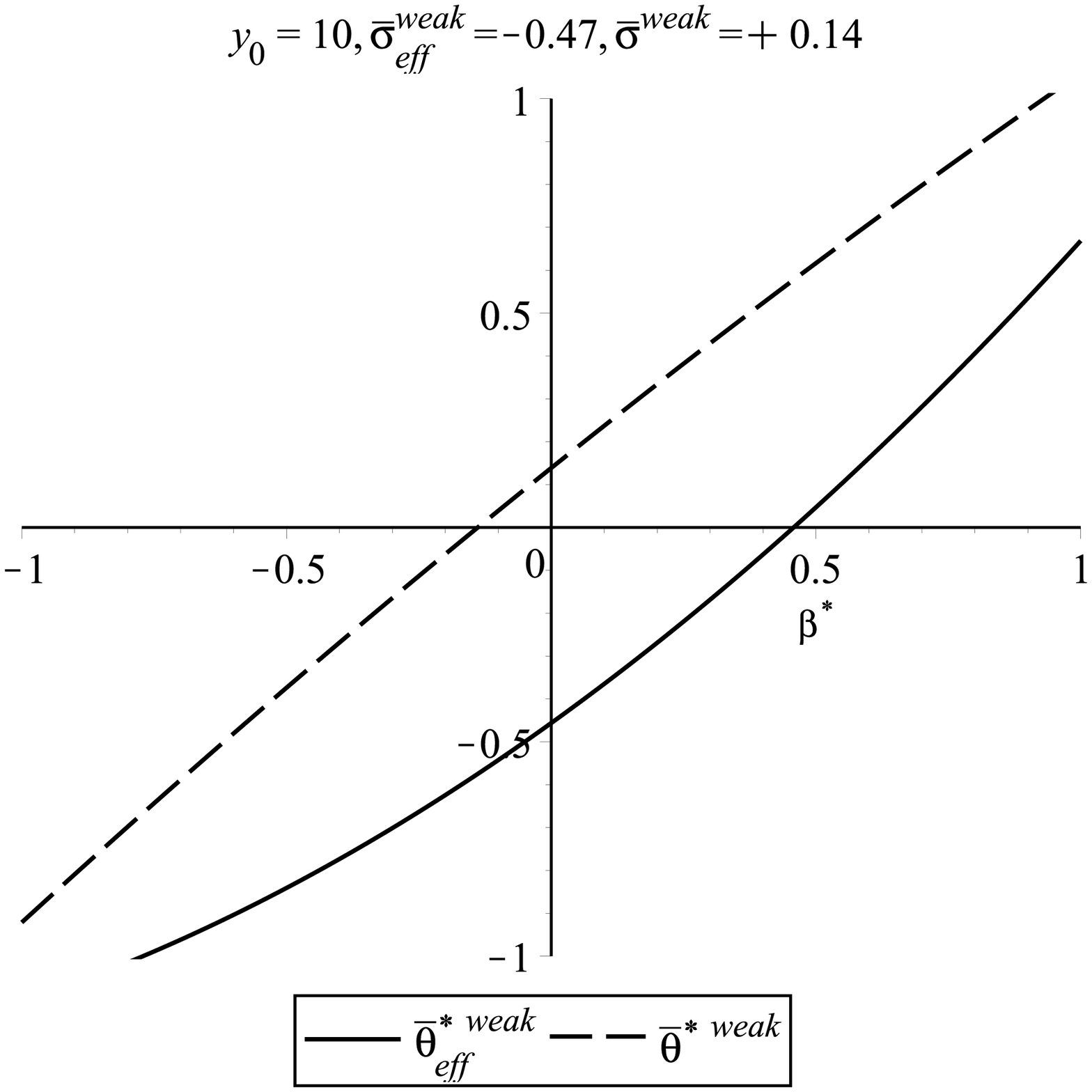}

\caption{{\small \ Diagrams of weak lensing primary image
locations for $|q|<1$ } }
\end{figure}
\begin{figure} \centering
\includegraphics[width=2.5in,height=2.2in]{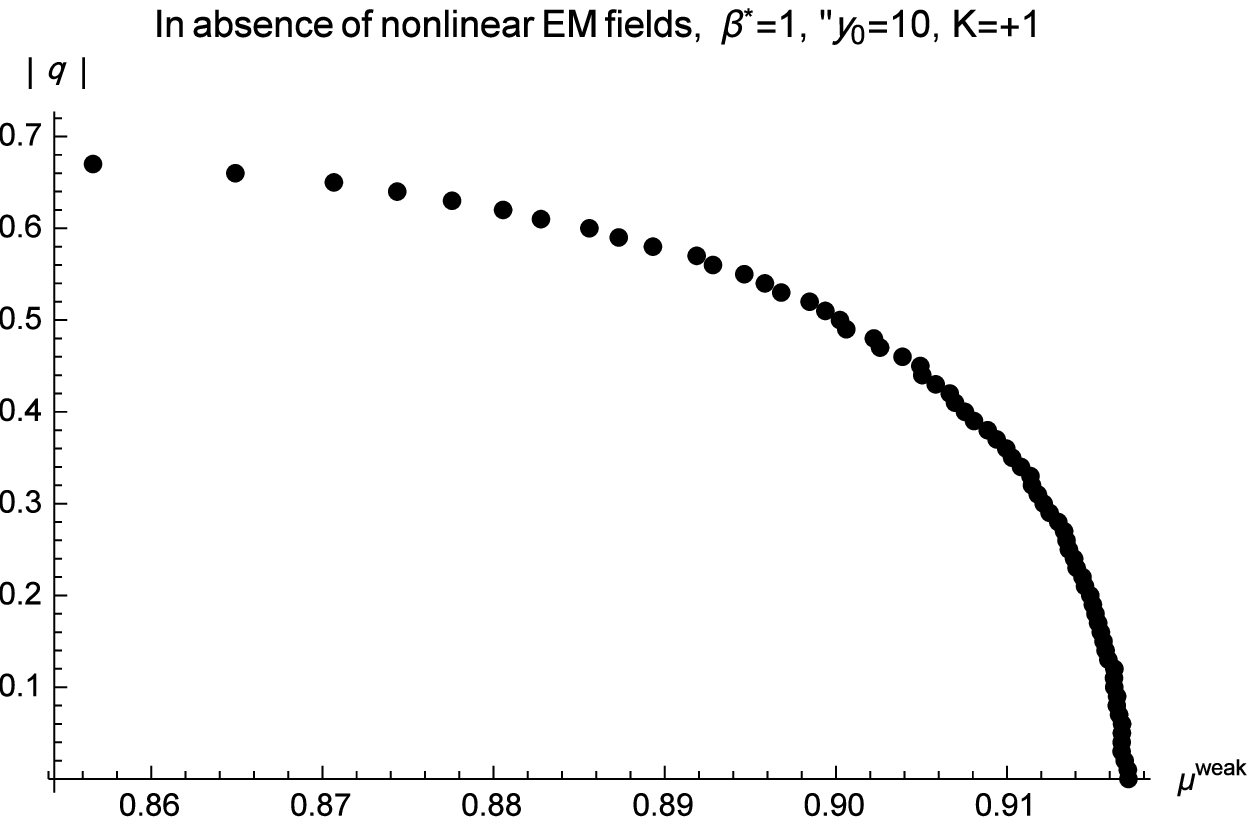}
\includegraphics[width=2.5in,height=2.2in]{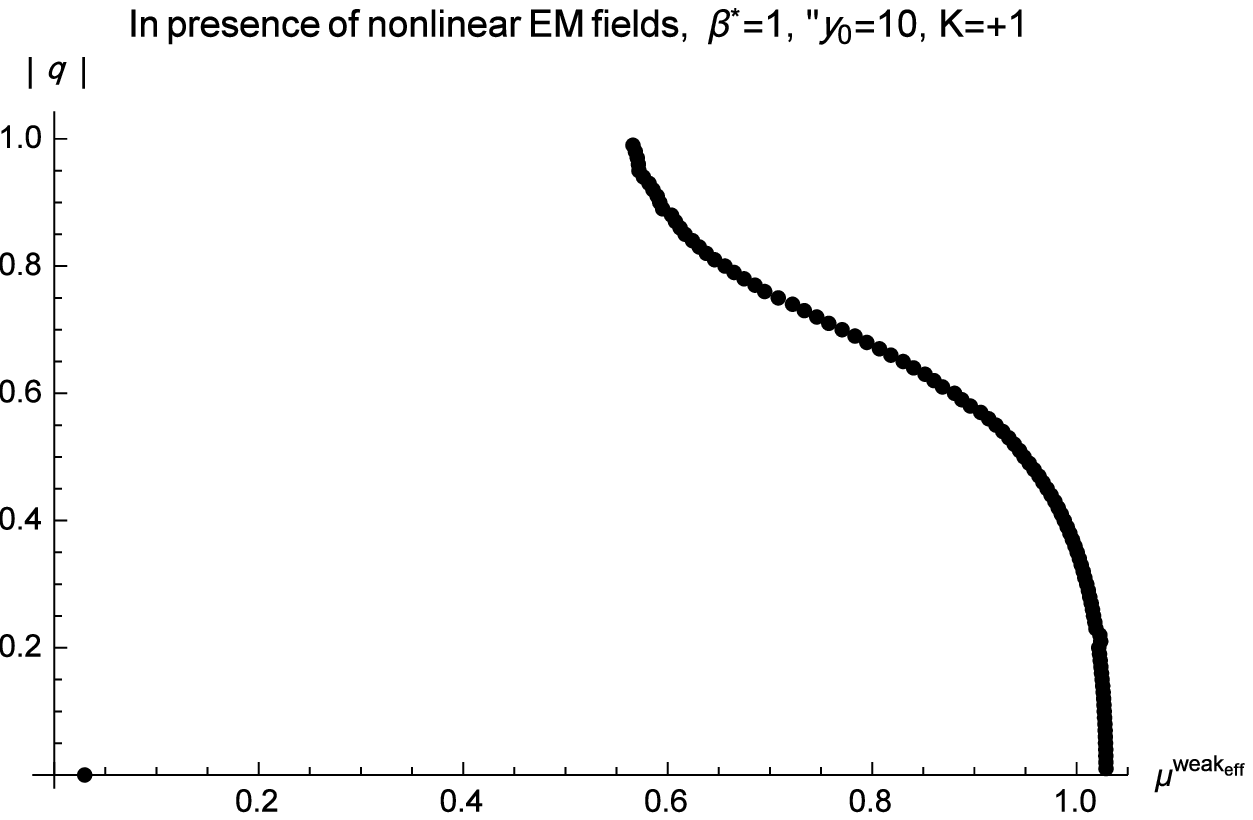}
\includegraphics[width=2.5in,height=2.2in]{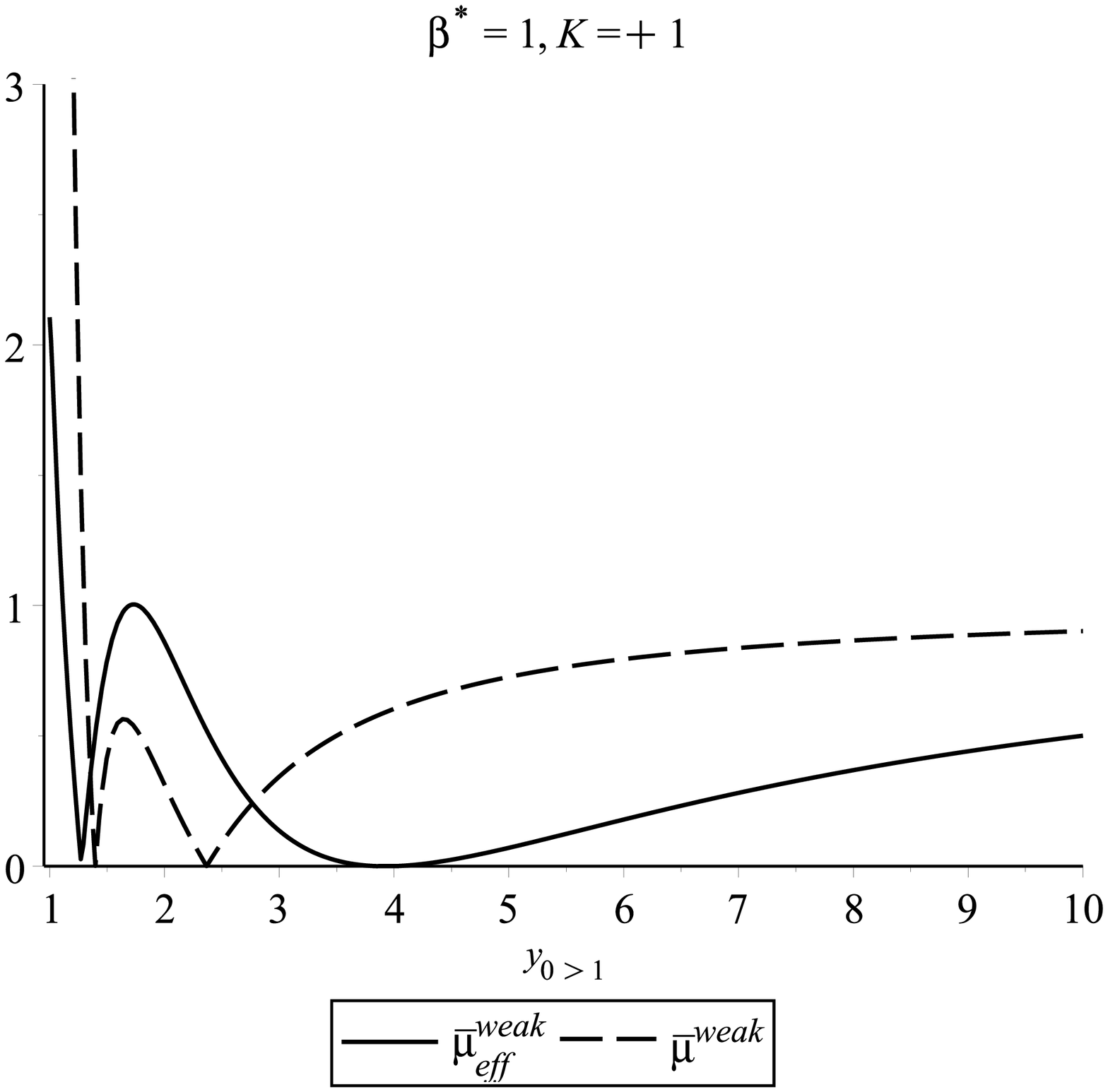}
\includegraphics[width=2.5in,height=2.2in]{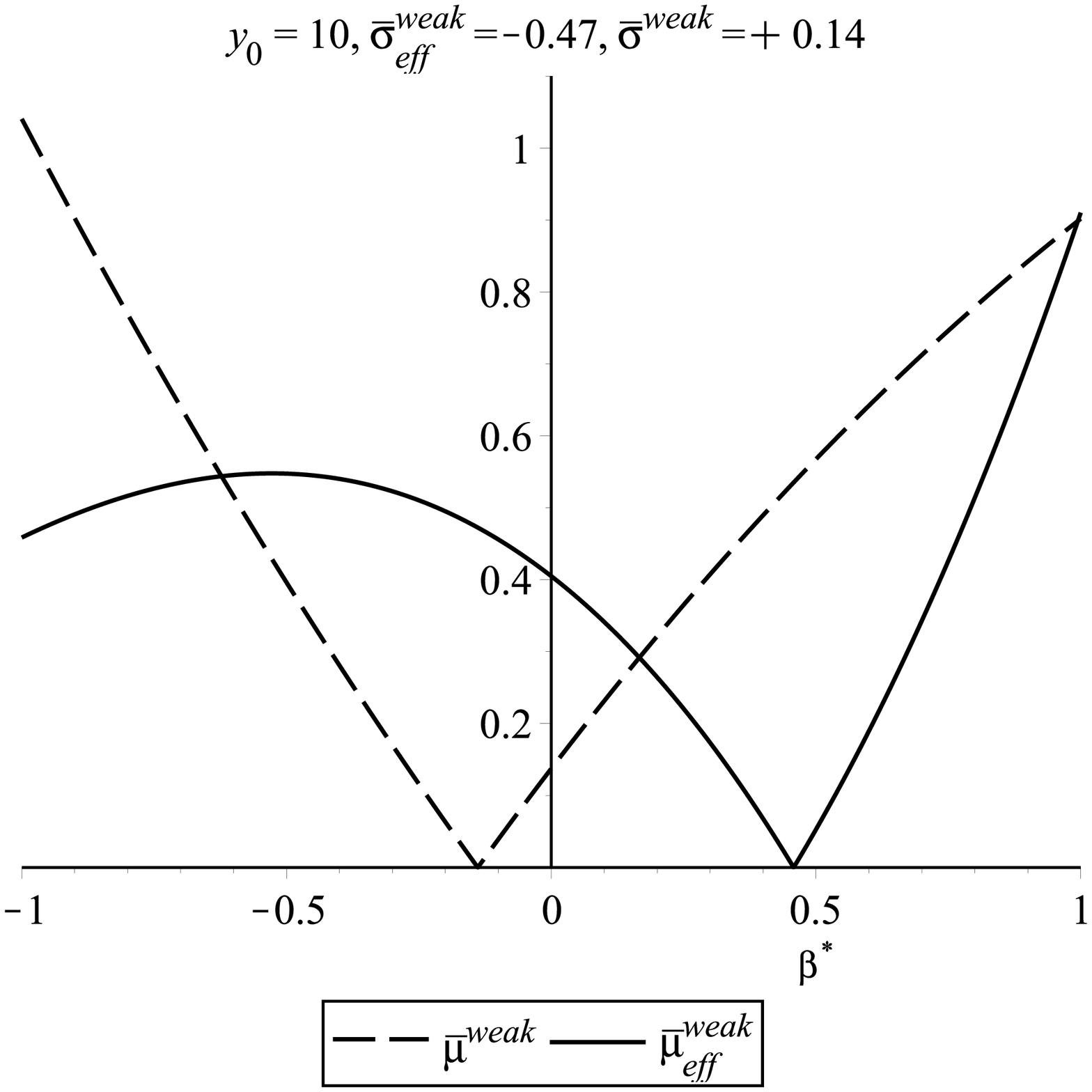}
\includegraphics[width=2.5in,height=2.2in]{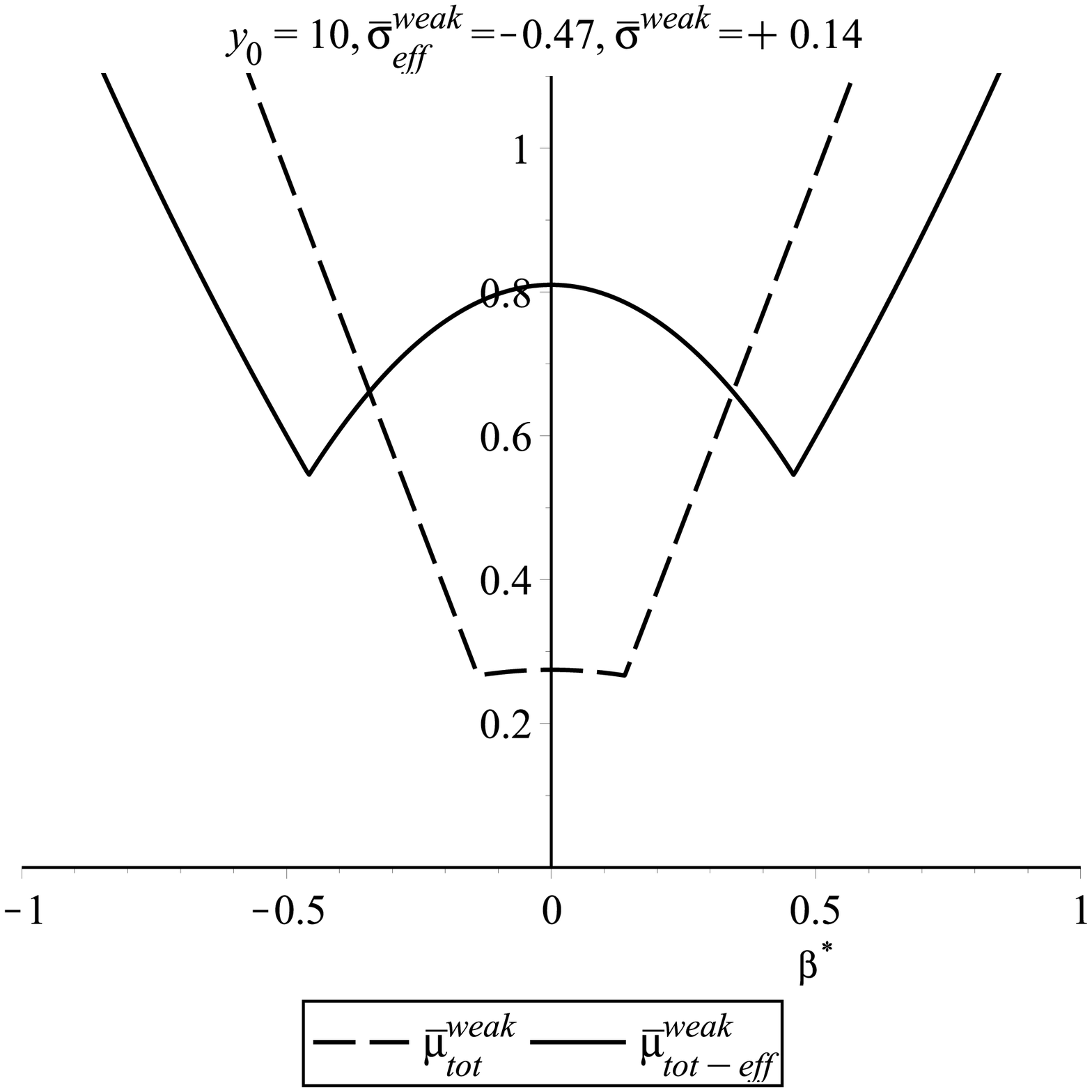}
\includegraphics[width=2.5in,height=2.2in]{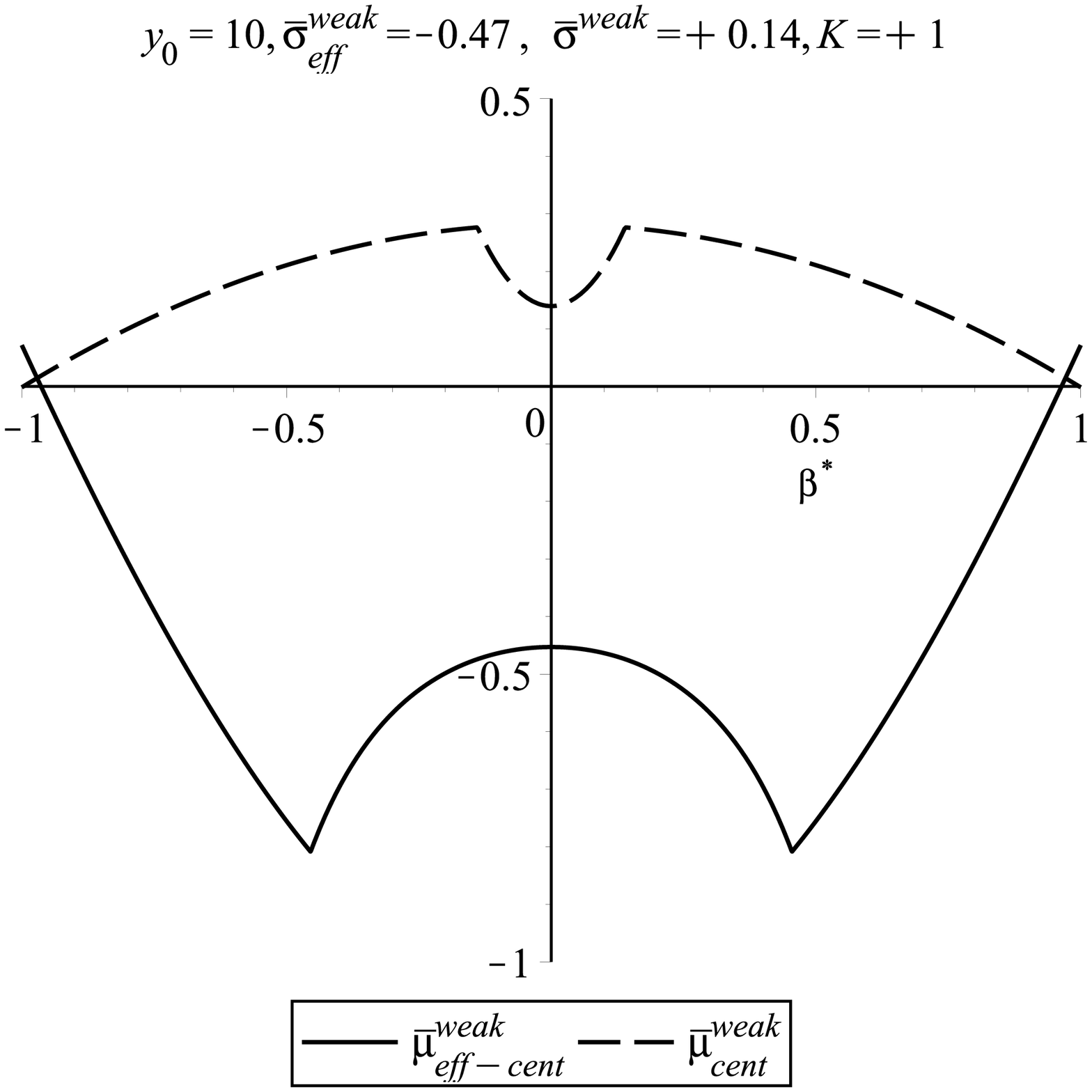}

\caption{{\small \ Diagrams of  weak lensing image magnifications
for $|q|<1$ } }
\end{figure}
\begin{figure} \centering
\includegraphics[width=2.5in,height=1.8in]{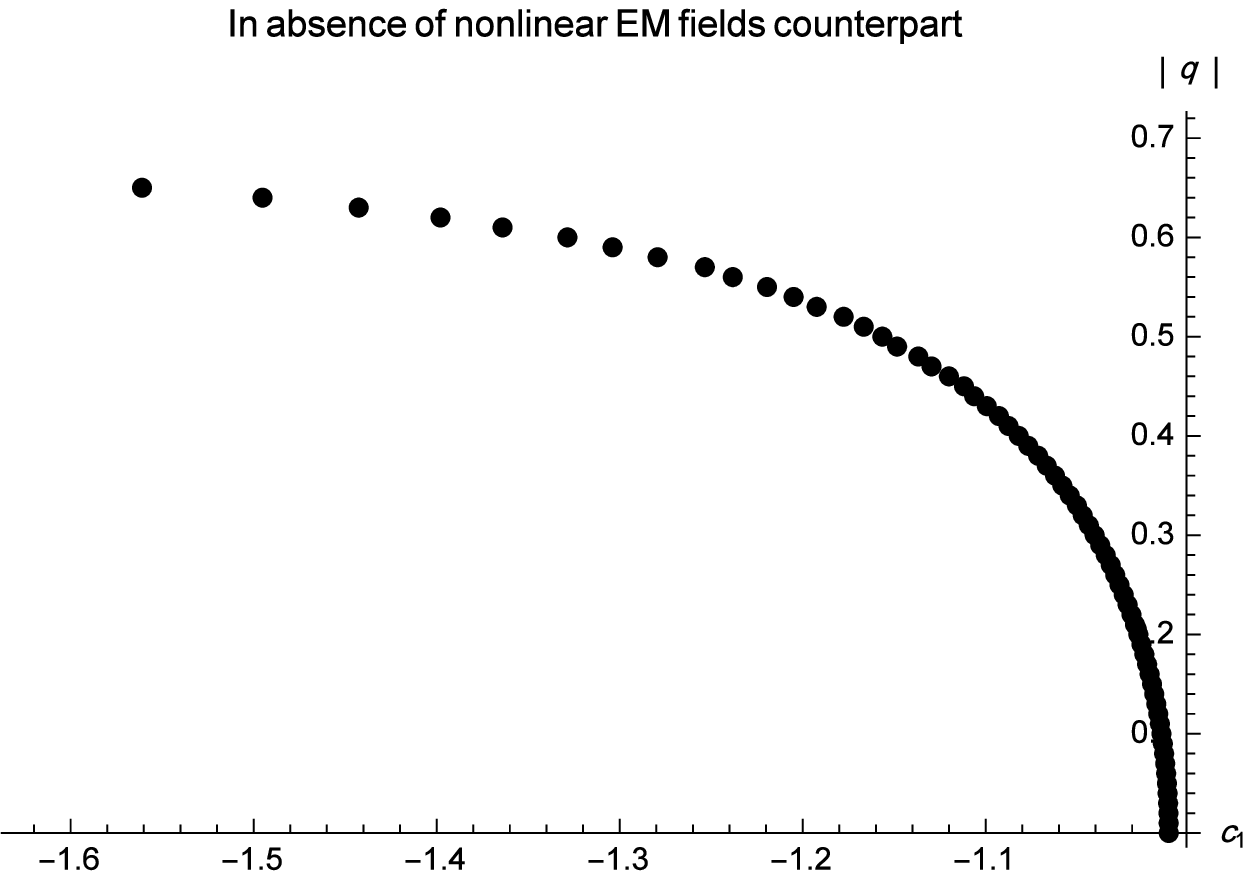}
\includegraphics[width=2.5in,height=1.8in]{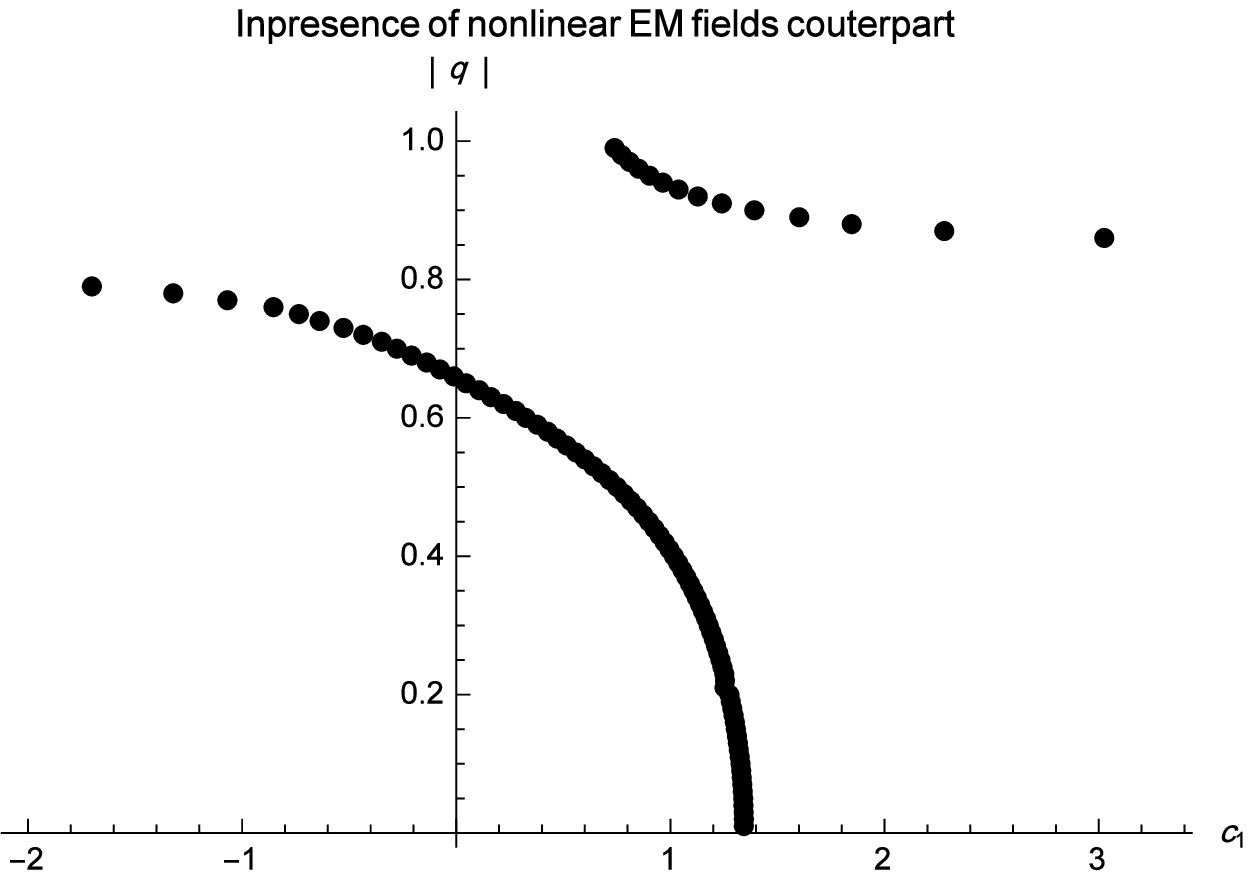}
\includegraphics[width=2.5in,height=1.8in]{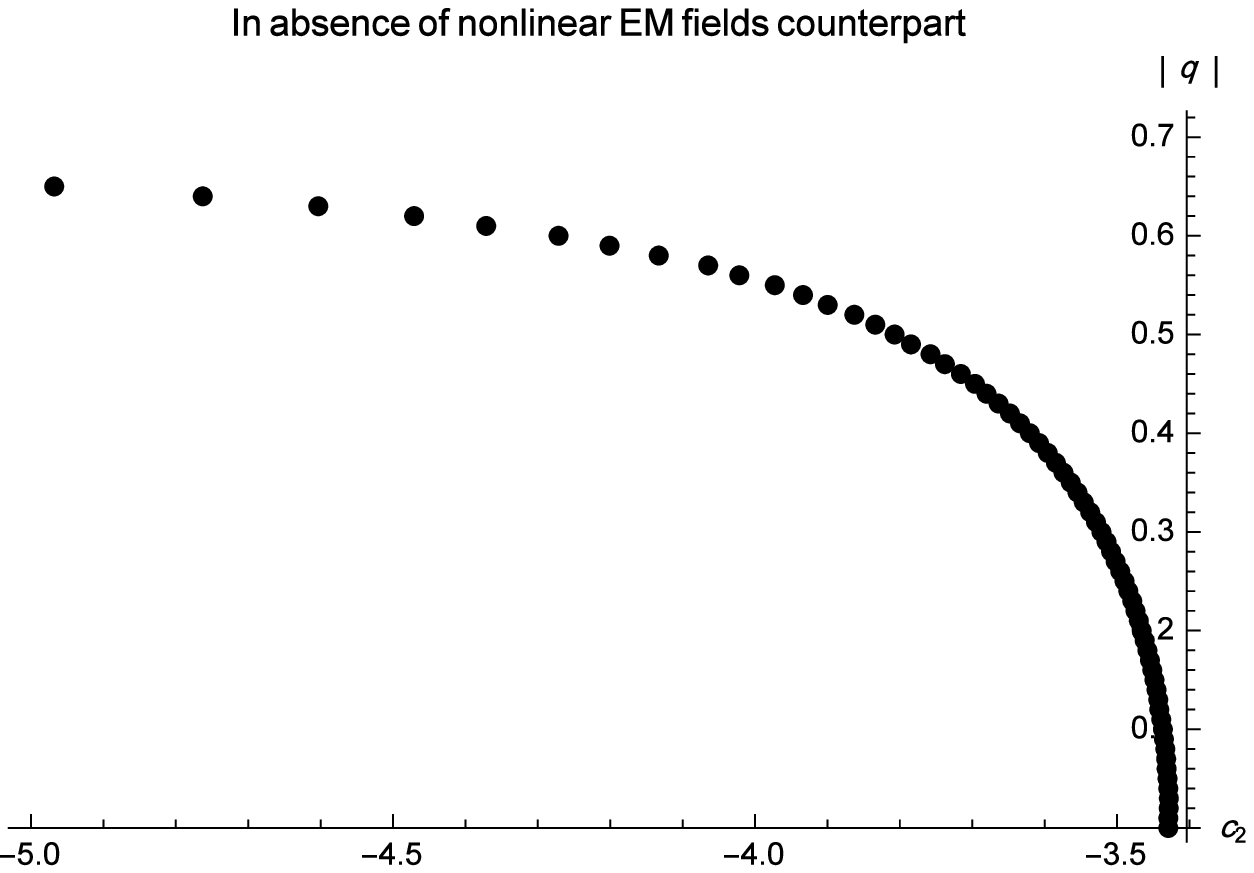}
\includegraphics[width=2.5in,height=1.8in]{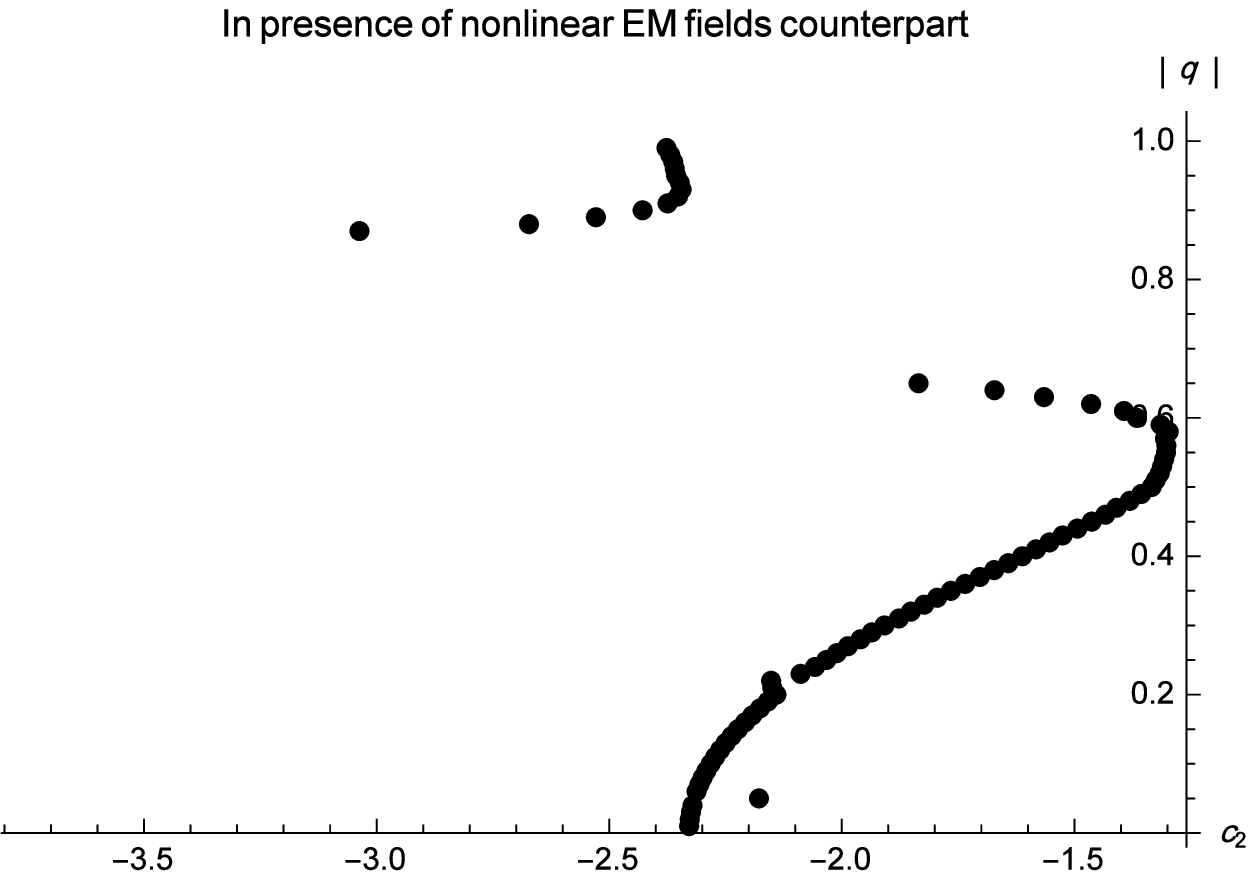}

\caption{{\small \ Diagrams of $c_1,c_2$ are plotted against
$|q|<1.$ }}
\end{figure}
\begin{figure} \centering
\includegraphics[width=2.5in,height=1.8in]{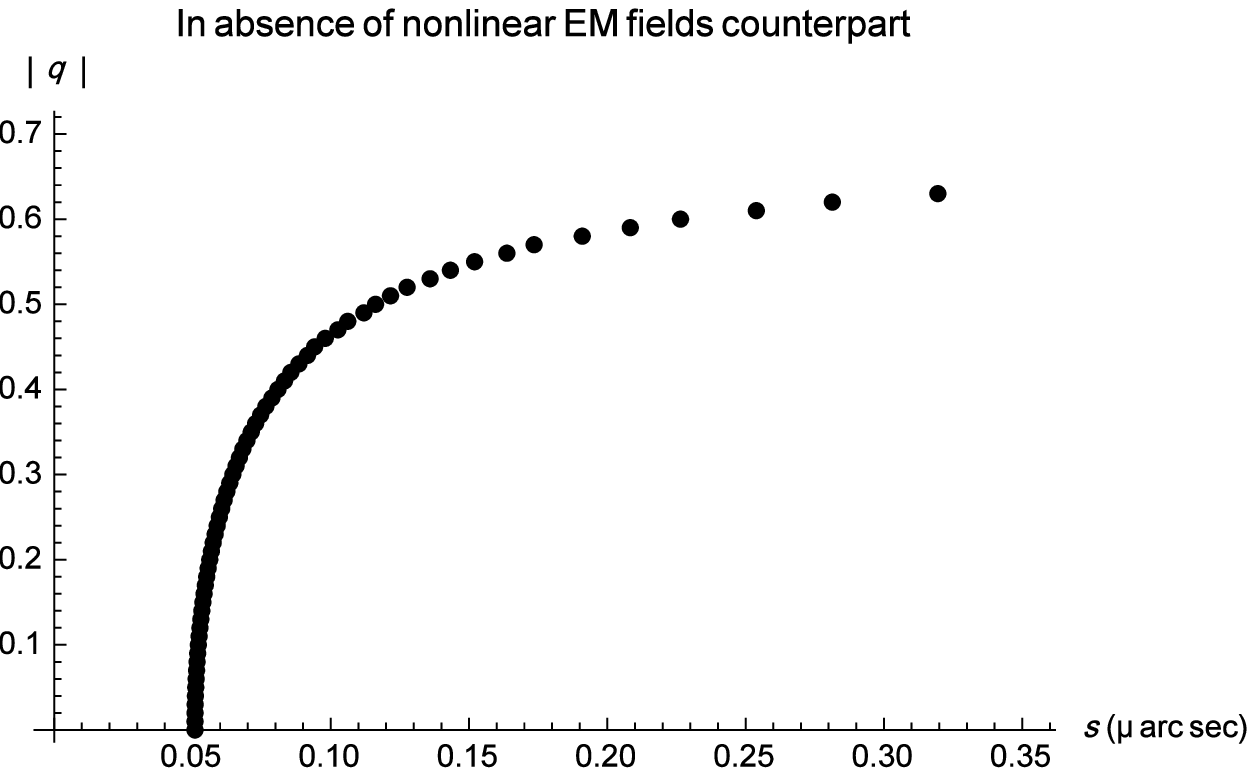}
\includegraphics[width=2.5in,height=1.8in]{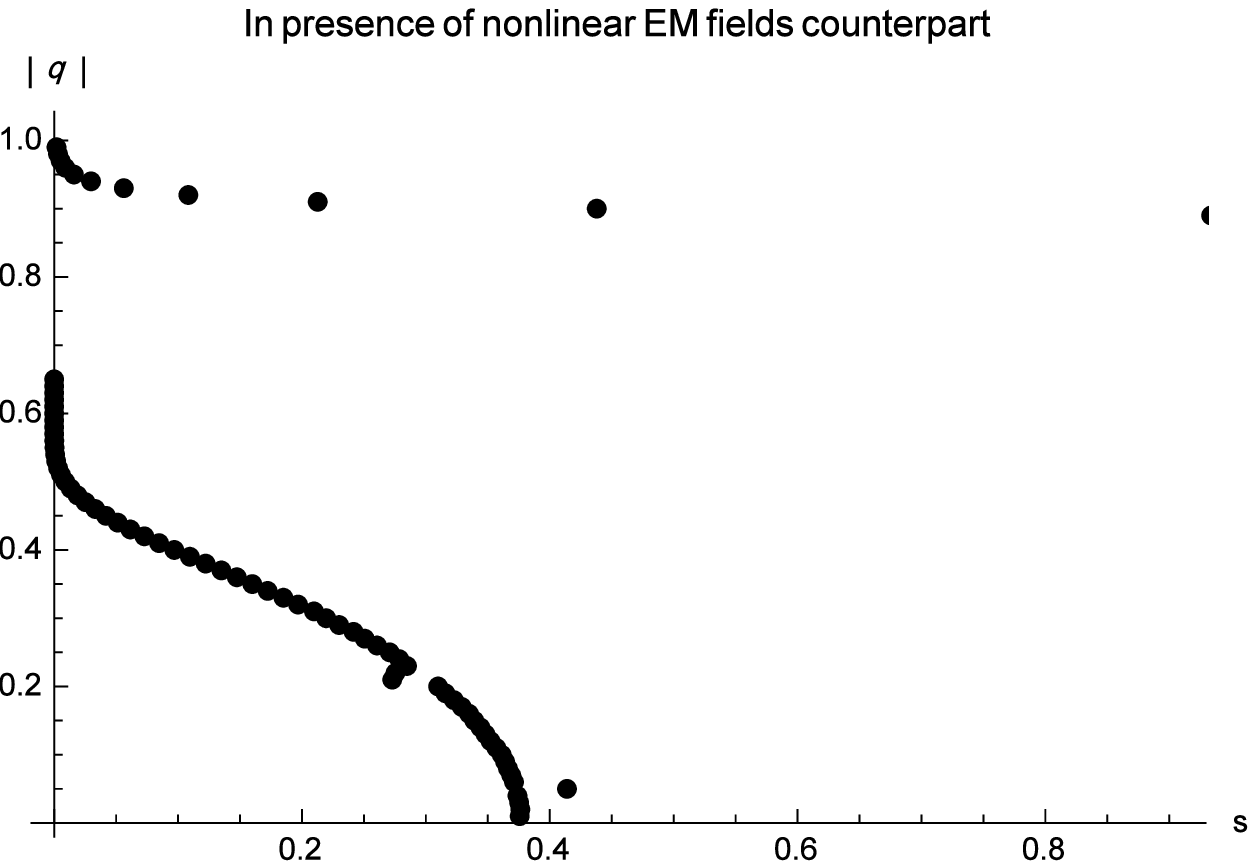}
\includegraphics[width=2.5in,height=1.8in]{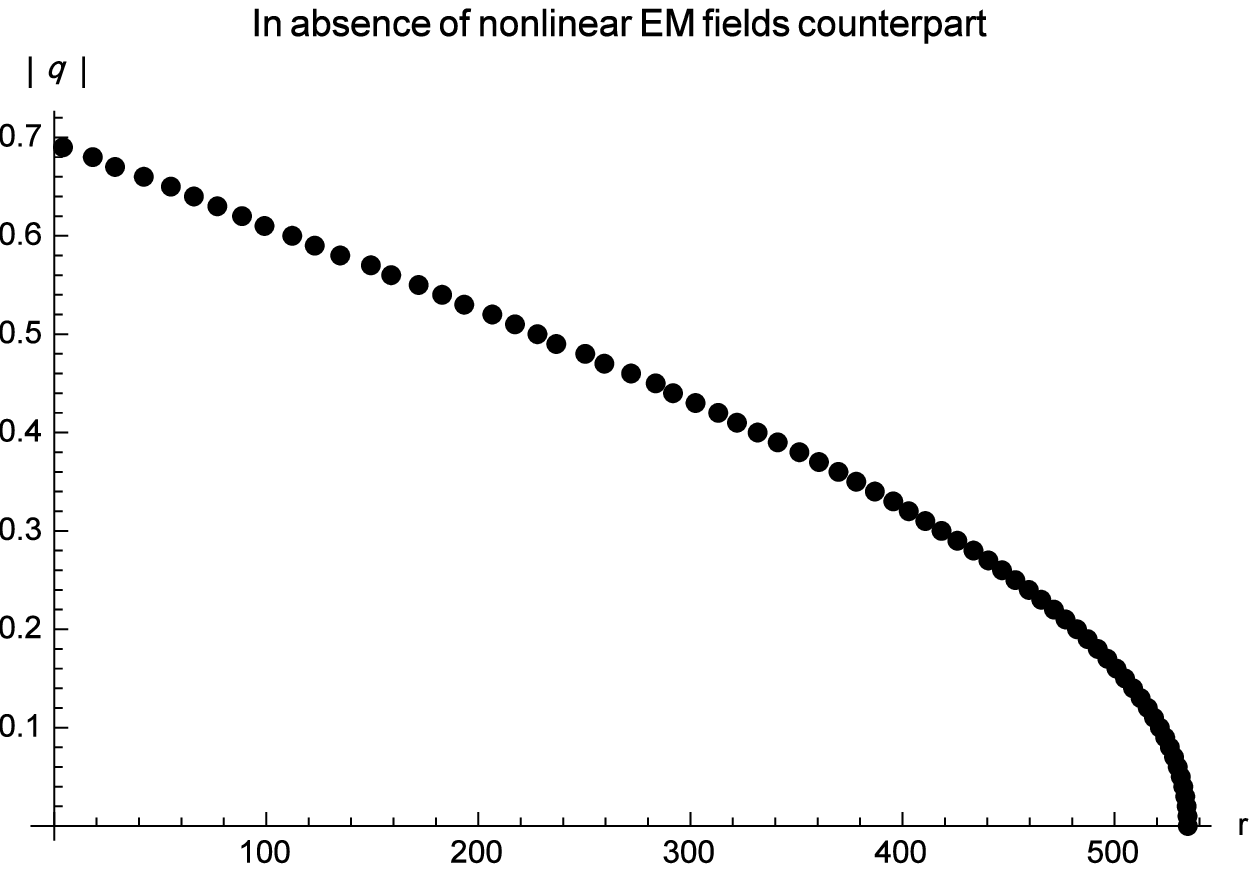}
\includegraphics[width=2.5in,height=1.8in]{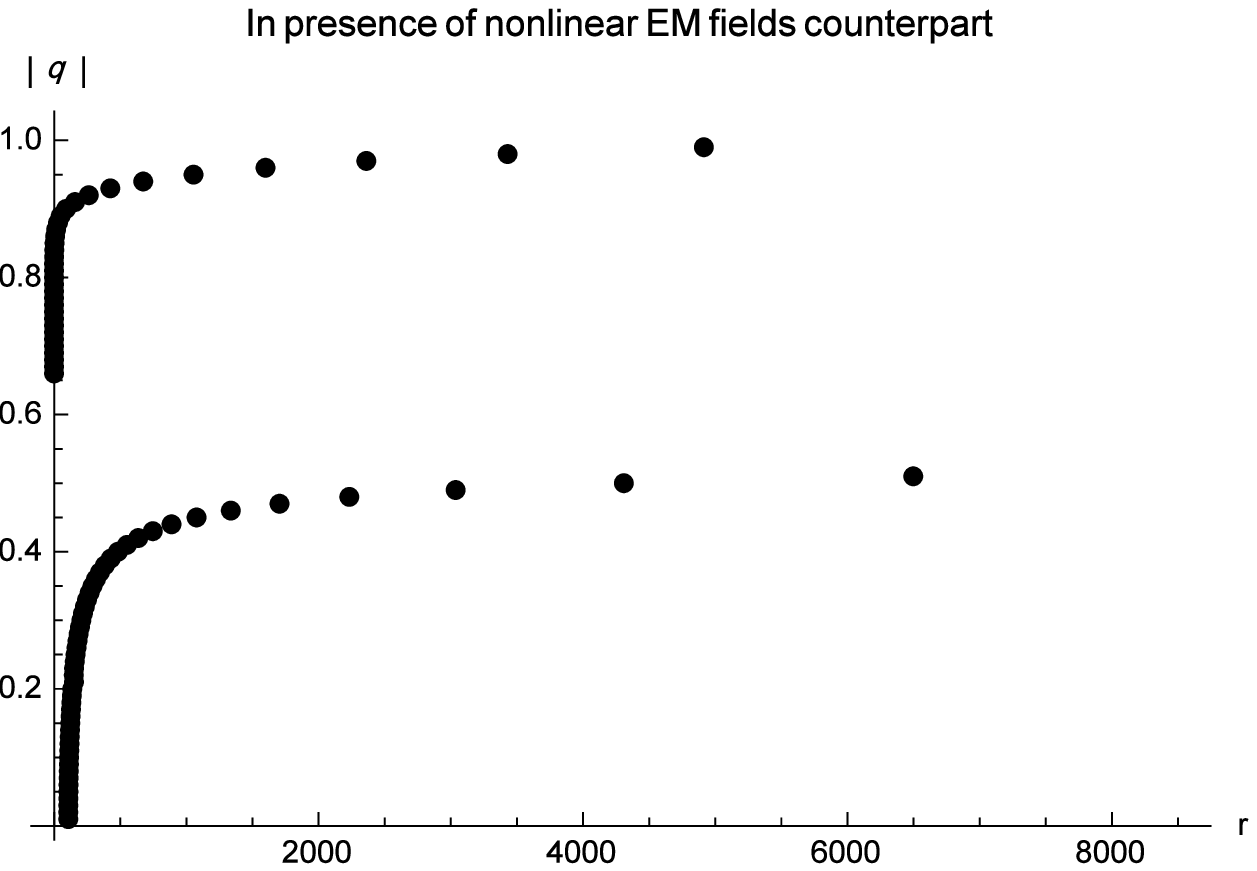}

\caption{{\small \ Diagrams of $s, r$ are plotted against $|q|<1.$
}}
\end{figure}
\begin{figure} \centering
\includegraphics[width=2.5in,height=2.2in]{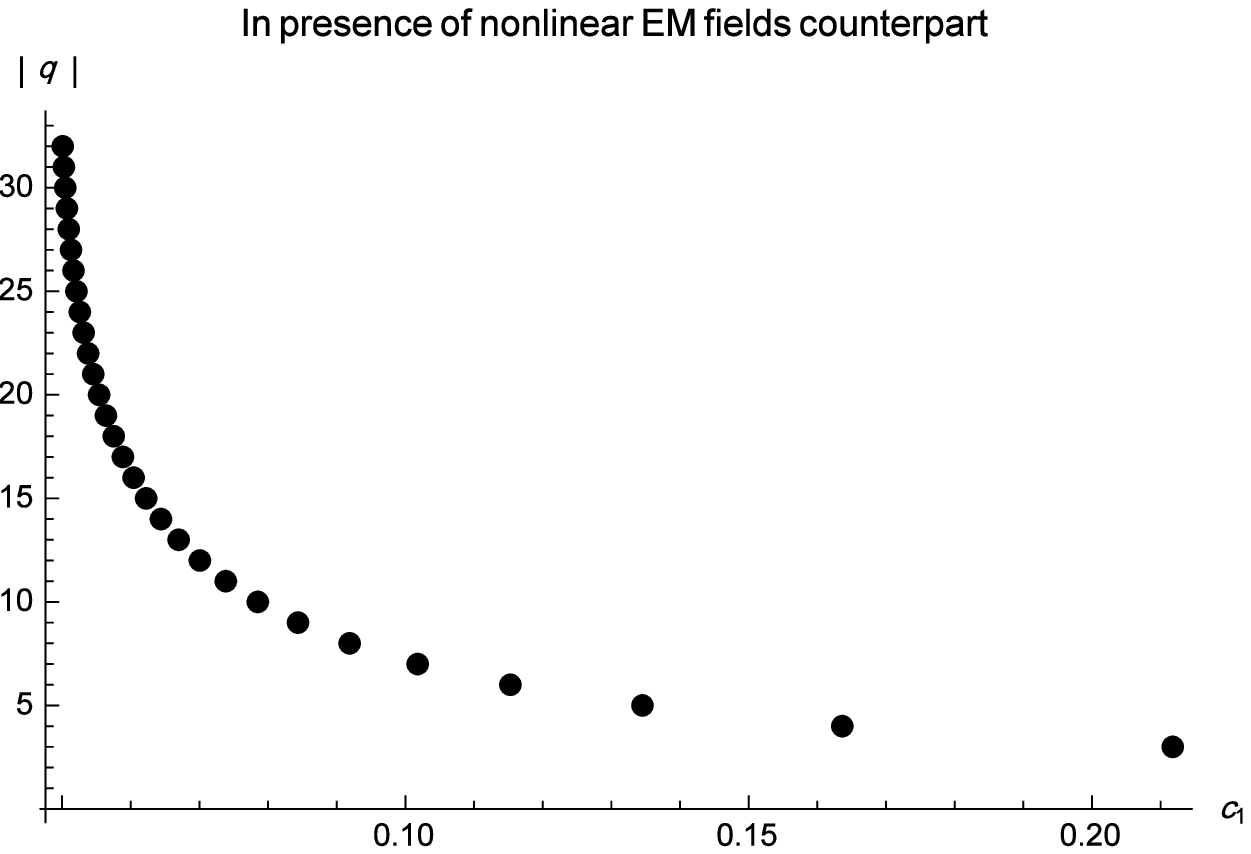}
\includegraphics[width=2.5in,height=2.2in]{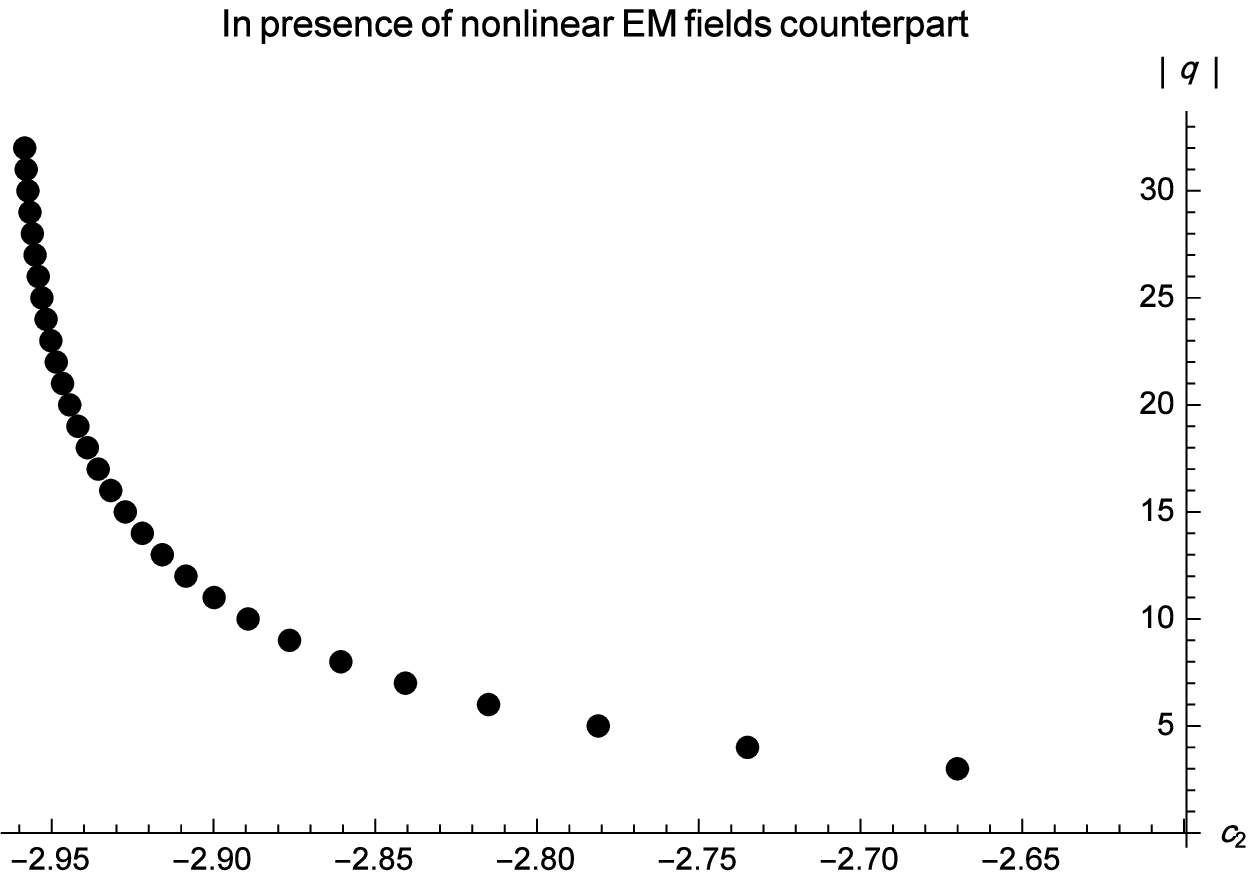}
\includegraphics[width=2.5in,height=2.2in]{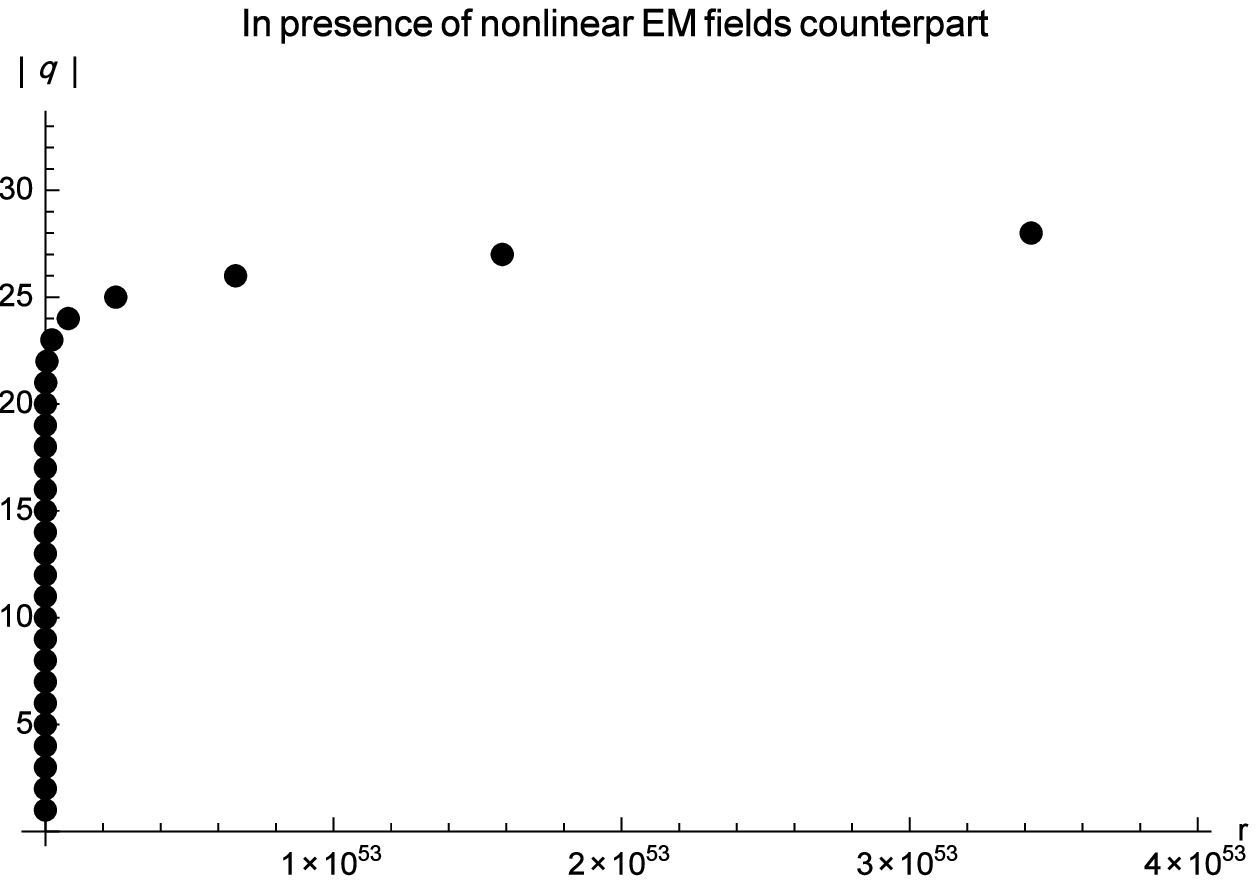}
\includegraphics[width=2.5in,height=2.2in]{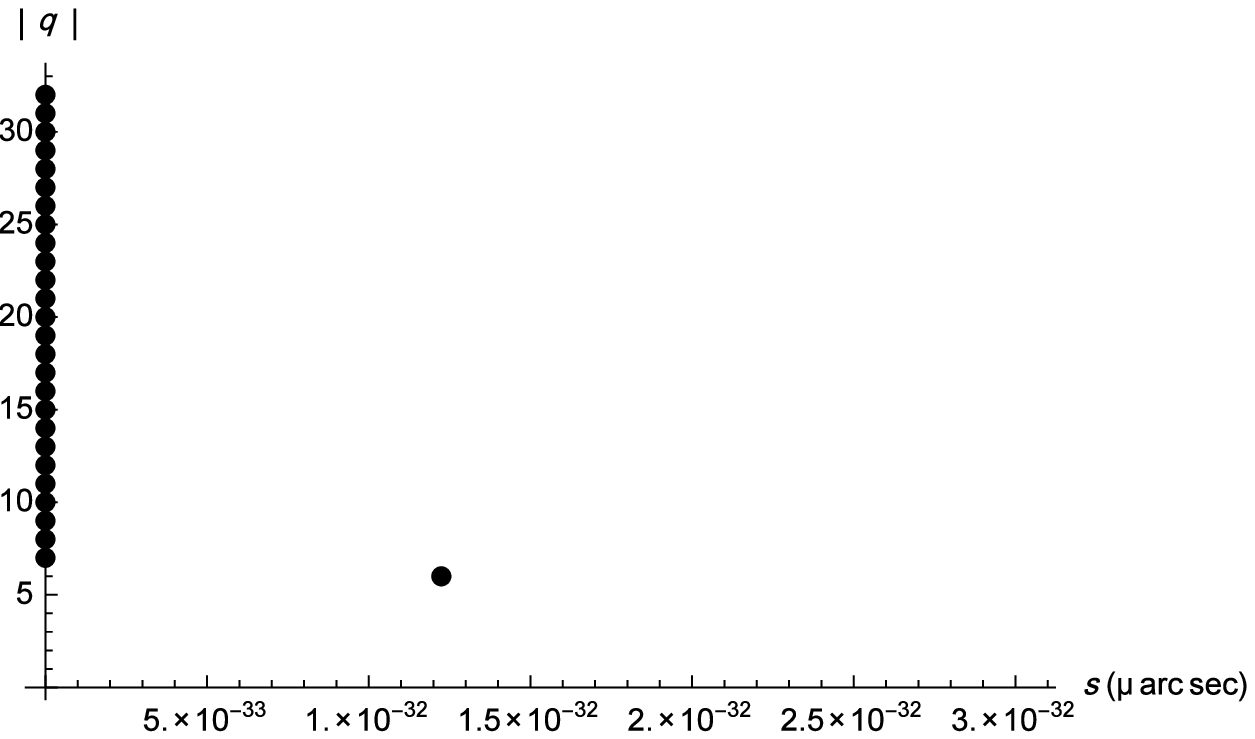}

\caption{{\small \ Diagrams of $c_1, c_2, s, r$ are plotted
against $|q|>1.$} }
\end{figure}

\newpage
\begin{center}
 Table 1. Numerical major real roots of the photon sphere
equations (3.17) and (3.19) for $0<|q|<1$.
\end{center}
\begin{center}
\begin{tabular}{|c|c|c|c|}
  \hline
$|q| ,$ $x_{ps}$, $x_{ps}^{eff}$ &$|q|$ , $x_{ps}$, $x_{ps}^{eff}$&$|q|$ , $x_{ps}$, $x_{ps}^{eff}$&$|q|$ , $x_{ps}$, $x_{ps}^{eff}$ \\
  \hline
   0.00, 3.000, 3.000 &0.25, 2.892,2.874&0.50, 2.547, 2.645& 0.75,-,2.326\\
  0.01, 2.999, 2.950 & 0.26, 2.887, 2.871&0.51, 2.528,2.640& 0.76, -,2.320\\
    0.02, 2.994, 2.947 &0.27, 2.882, 2.865&0.52, 2.508, 2.630& 0.77, -,2.326\\
 0.03, 2.989, 2.947 &0.28, 2.872, 2.855&0.53, 2.475, 2.621& 0.78, -,2.329\\
0.04, 2.989, 2.947  &0.29, 2.858, 2.852&0.54, 2.455, 2.609& 0.79, -,2.335\\
 0.05, 2.989, 2.944 &0.30, 2.848, 2.843 &0.55, 2.431, 2.594& 0.80, -,2.344\\
  0.06, 2.989, 2.943 &0.31, 2.838, 2.831&0.56, 2.397,2.578& 0.81, -,2.350\\
  0.07, 2.984, 2.942 &0.32, 2.828, 2.828&0.57, 2.378,2.560& 0.82, -,2.362\\
  0.08,2.980, 2.941 &0.33, 2.824, 2.819&0.58, 2.334, 2.530& 0.83, -,2.377\\
  0.09, 2.980, 2.938 &0.34, 2.809, 2.813&0.59, 2.300,2.514& 0.84, -,2.393\\
 0.10, 2.975, 2.935 &0.35, 2.795,2.804&0.60, 2.271,2.508& 0.85, -,2.408\\
 0.11, 2.974, 2.935&0.36, 2.785, 2.795&0.61, 2.227,2.478& 0.86, -,2.429\\
  0.12, 2.974, 2.935 &0.37, 2.770,2.785&0.62, 2.193,2.469& 0.87, -,2.451\\
  0.13, 2.964, 2.932 &0.38, 2.756, 2.776&0.63, 2.150,2.460& 0.88, -,2.475\\
    0.14, 2.959, 2.928 &0.39, 2.736, 2.767&0.64, 2.106,2.441& 0.89, -,2.487\\
 0.15, 2.955, 2.925 &0.40, 2.722, 2.758&0.65, 2.058,2.429& 0.90, -,2.514\\
 0.16, 2.950, 2.919  &0.41, 2.707,2.749&0.66, 1.990, 2.411& 0.91, -,2.542\\
 0.17, 2.945, 2.916 &0.42, 2.698, 2.740 &0.67, 1.902,2.399&0.92, -,2.566 \\
  0.18, 2.940, 2.913 &0.43, 2.678, 2.731&0.68, 1.819,2.384& 0.93, -,2.591\\
 0.19, 2.935, 2.910 &0.44, 2.659,2.718&0.69, 1.645, 2.374& 0.94, -,2.612\\
  0.20, 2.930, 2.904 &0.45, 2.654, 2.706&0.70, -,2.362& 0.95, -,2.636\\
  0.21, 2.921, 2.989 &0.46, 2.630,2.694&0.71, -,2.350& 0.96, -,2.670\\
 0.22, 2.916,2.895 &0.47, 2.601, 2.685&0.72, -,2.344&0.97, -,2.703 \\
 0.23, 2.906, 2.889&0.48, 2.591,2.670&0.73, -,2.338&0.98, -,2.734\\
  0.24, 2.901, 2.883 &0.49, 2.557,2.658&0.74, -,2.335&0.99,-,2.764 \\
  \hline
\end{tabular}
\end{center}
\begin{center}
 Table 2.   Solutions of the effective photon sphere
equation (3.20) for $|q|>1$.
\end{center}
\begin{tabular}{|c|c|c|c|c|c|}
  \hline
  $|q|,x^{eff}_{ps}$ &  $|q|,x^{eff}_{ps}$ &  $|q|,x^{eff}_{ps}$ &  $|q|,x^{eff}_{ps}$&  $|q|,x^{eff}_{ps}$ &  $|q|,x^{eff}_{ps}$ \\
  \hline
1, 2.847 & 7, 24.705 & 13,46.563 & 19, 68.421& 25, 90.279 & 31, 112.137\\
 2, 6.490 & 8, 28.348 & 14,50.206 & 20, 72.064& 26, 93.922& 32, 115.780\\
  3, 10.133 & 9, 31.991& 15,53.849 & 21, 75.707 &27, 97.565 &33, 119.423\\
  4, 13.776 & 10, 35.634& 16,57.492 & 22, 79.350& 28, 101.208& 34, 123.066\\
  5, 17.419 & 11, 39.277 & 17, 61.135& 23,82.993 & 29, 104.851& 35, 126.709\\
  6,  21.062 & 12, 42.920& 18,64.778 & 24,86.636 & 30, 108.494& 36, 130.352\\
  \hline
\end{tabular}
\end{document}